\documentclass[1p]{elsarticle}
\usepackage{soul}
\usepackage{lineno}
\usepackage{hyperref}
\usepackage{amsmath}
\usepackage{graphicx}
\usepackage{subfig}
\usepackage{array}
\usepackage{amsfonts}
\usepackage[dvipsnames]{xcolor}
\usepackage{booktabs}
\modulolinenumbers[10]

\journal{arXiv} 

\begin{document}

\begin{frontmatter}

\title{Diagnosing numerical Cherenkov instabilities in relativistic plasma simulations based on general meshes}

\author[ad1]{D.-Y. Na}
\ead{na.94@osu.edu}
\author[ad1]{~J. L. Nicolini}
\ead{delimanicolini.1@osu.edu}
\author[ad1]{~R. Lee}
\ead{lee.146@osu.edu}
\author[ad2]{~B.-H. V. Borges}
\ead{benhur@sc.usp.br}
\author[ad3]{~Y. A. Omelchenko}
\ead{omelche@gmail.com} 
\author[ad1]{F. L. Teixeira}
\ead{teixeira.5@osu.edu}

\address[ad1]{ElectroScience Laboratory and Department of Electrical and Computer Engineering, The Ohio State University, Columbus OH, 43212-1166, USA}
\address[ad2]{Department of Electrical and Computer Engineering, University of S\~{a}o Paulo, S\~{a}o Carlos SP, 13566-590, Brazil}
\address[ad3]{Trinum Research Inc., San Diego CA, 92126-5587, USA }

\begin{abstract}
Numerical Cherenkov radiation (NCR) or instability is a detrimental effect frequently found in electromagnetic particle-in-cell (EM-PIC) simulations involving relativistic plasma beams. NCR is caused by spurious coupling
between
electromagnetic-field modes and
multiple beam resonances. This coupling may result from the slow down of poorly-resolved waves due to numerical (grid) dispersion and from aliasing mechanisms.
NCR has been studied in the past for finite-difference-based EM-PIC algorithms on
regular (structured) meshes with rectangular elements.
In this work, we extend the analysis of NCR to finite-element-based EM-PIC algorithms implemented on unstructured meshes.
The influence of different mesh element shapes and mesh layouts on NCR is studied.
Analytic predictions are compared against results from finite-element-based EM-PIC simulations of relativistic plasma beams on various mesh types.
\end{abstract}

\begin{keyword}
Particle-in-cell, numerical Cherenkov radiation, finite element method, plasmas.
\end{keyword}

\end{frontmatter}


\section{Introduction}\label{sec:Introduction}
Electromagnetic particle-in-cell (EM-PIC) simulations have become an important tool for the study of a wide variety of problems associated with collisionless plasmas, including high power vacuum electronic devices~\cite{gold1997review,cairns1997generation,na2017axisymmetric}, laser-wakefield acceleration~\cite{esarey2009physics}, and astrophysical phenomena~\cite{matsumoto2013electron}, to name just a few.
Despite their success, there exist a number of outstanding challenges that limit the accuracy and robustness of EM-PIC simulations. Among them, 
the numerical Cherenkov radiation (NCR) instability, first observed by Godfrey~\cite{godfrey1974numerical}, has been recognized as an important detrimental factor in EM-PIC simulations involving high-energy (relativistic) charged particles (including Lorentz-boosted frames)~\cite{vay2011numerical,lehe2013numerical} and collisionless shocks \cite{bret2013collisionless}. 
On regular periodic meshes such as used by the finite-difference time-domain (FDTD) method, NCR results from the coupling between numerical 
electromagnetic modes and
plasma beam resonances.  This coupling may result from the slow down of poorly-resolved waves due to numerical dispersion or 
from aliasing mechanisms~\cite{godfrey2013numerical,xu2013numerical}.

The study of the causes and behavior of NCR is of critical importance for developing mitigation strategies~\cite{godfrey2013numerical,xu2013numerical,greenwood2004elimination,godfrey2014numerical,godfrey2015improved,vay2013domain}.
NCR has been extensively studied for FDTD-based and spectral-based EM-PIC algorithms based on regular meshes~\cite{godfrey1974numerical,vay2011numerical,lehe2013numerical,bret2013collisionless,godfrey2013numerical,xu2013numerical,greenwood2004elimination}. 
For problems involving complex geometries, however, it is often advantageous to employ more general meshes which can better conform to curved and/or irregular boundaries as well as support adaptive mesh refinement capabilities.

In this work, we analyze NCR effects in EM-PIC simulations based on more general meshes.
The finite element time-domain (FETD)-based EM-PIC algorithm discussed in~\cite{moon2015exact,na2016local,na2018relativistic} is employed for this purpose. This explicit algorithm includes a charge-conserving scatter algorithm~\cite{moon2015exact} and the Higuera-Cary particle-pusher to fully take into account relativistic effects and yield an overall energy-conserving algorithm~\cite{na2018relativistic,cary2017structure}. 
{\color{black}The present EM-PIC algorithm achieves discrete energy conservation bounded by finite precision errors since the FETD field solver is symplectic and the Higuera-Cary particle pusher is volume-preserving in the phase space.} The reason for adopting this FETD-based EM-PIC algorithm in this study is twofold: (1) Contrary to most past FE-based EM-PIC algorithms implemented on general meshes~\footnote{Some notable exceptions are the compatible FE-based formulations described in Refs.~\cite{squire2012geometric,pinto2014charge,kraus_2017}}, the present algorithm attains both charge and energy-conservation from first principles. (2)  The standard FDTD algorithm can be retrieved as a special version of this mixed FETD scheme implemented on a regular mesh with square elements, in which low-order quadrature rules are employed in the evaluation of the mass (Hodge) matrices elements to yield diagonal matrices and a fully explicit field update~\cite{lee2006note}. 
 This facilitates a direct comparison of NCR effects arising in FETD-based EM-PIC simulations with those in FDTD-based EM-PIC simulations.

As noted, NCR is closely related to numerical dispersion. Most past numerical dispersion studies in FE-based Maxwell field solvers have been carried out in the frequency domain (or equivalently, with no time discretization errors included) \cite{wu1997advantages,lee1992study,warren1994investigation} and only a few in the time domain~\cite{monorchio2002dispersion}. 
Key conclusions from these studies are as follows: (1) a good quality mesh (i.e. one having near equilateral elements) is best for minimizing {\it local} phase errors per wavelength
and (2) the {\it cumulative} phase error can be smaller on highly unstructured grids due to cancellation effects.
The numerical dispersion analysis carried out in these works have focused on well-resolved waves, which is the practical regime of interest to provide sufficiently accurate results for pure EM simulations. However, in order to analyze NCR in EM-PIC simulations, a complete numerical dispersion map over the first Brillouin zone (the periodic layout of the mesh elements) must be determined because, among other reasons, electric currents, mapped from moving charged particles to the mesh during the scatter step of EM-PIC algorithm, are not explicitly bandlimited in contrast to the typical sources present in particle-free EM simulations. In this work, complete dispersion diagrams over the first Brillouin zone are obtained for meshes with different element shapes and layouts and analytical predictions for NCR are compared with numerical results from EM-PIC simulations.

The remainder of this paper is organized as follows.
To facilitate comparison, NCR effects that arise from FDTD-based EM-PIC algorithms on regular grids are summarized in Sec.~\ref{sec:FDTD}.
NCR effects that arise from FETD-based EM-PIC algorithms on general grids are analyzed in Sec.~\ref{sec:FETD_Dispersion}
and contrasted with the results found in Section 2. A systematic approach to deriving numerical dispersion curves is employed to study the influence of different mesh element shapes and mesh layouts on NCR properties.
In Sec.~\ref{sec:Numerical_Experiment}, EM-PIC simulations of relativistically drifting plasma beams are presented to validate the analytic predictions made 
in Sec.~\ref{sec:FETD_Dispersion}.
Some concluding remarks are provided in Sec.~\ref{sec:Conclusion}.

\section{Numerical Cherenkov Radiation in the FDTD-based EM-PIC Algorithm}\label{sec:FDTD}
Owing to its flexibility and robustness, the FDTD algorithm is arguably the most popular field solver for time-domain Maxwell's equations~\cite{taflove2000computational}. As such, FDTD-based (Yee) EM-PIC simulations are widely employed in plasma physics applications. The FDTD uses central-difference approximations for both space and time derivatives applied on a structured regular mesh. 
The numerical dispersion for the FDTD algorithm in 2-D takes the form~\cite{taflove2000computational}
\begin{flalign}
\left[\frac{1}{h}\sin{\left(\frac{\tilde{\kappa}_x h}{2}\right)}\right]^{2}+\left[\frac{1}{h}\sin{\left(\frac{\tilde{\kappa}_y h}{2}\right)}\right]^{2}
=
\left[\frac{1}{c\Delta t}\sin{\left(\frac{\omega \Delta t}{2}\right)}\right]^{2},
\label{eqn:gd_fdtd}
\end{flalign}
where $c$ is the speed of light in vacuum, $\Delta t$ is the time step interval, $h$ is the edge length of a square unit cell in the structured mesh, and $\tilde{\boldsymbol{\kappa}}=\tilde{\kappa}_{x}\hat{x}+\tilde{\kappa}_{y}\hat{y}$ is the 2-D wavenumber for plane waves propagating on the mesh.
We use the tilde to indicate numerical wavenumber $\tilde{\boldsymbol{\kappa}}$ (as modified by numerical dispersion) as opposed to exact wavenumber
$\boldsymbol{\kappa}$. Throughout this work, the time convention $e^{j\omega t}$ is adopted.

\begin{figure}
\centering
\subfloat[\label{fig:gd_fdtd_cfl_1_3d}]{\includegraphics[width=2.75in]{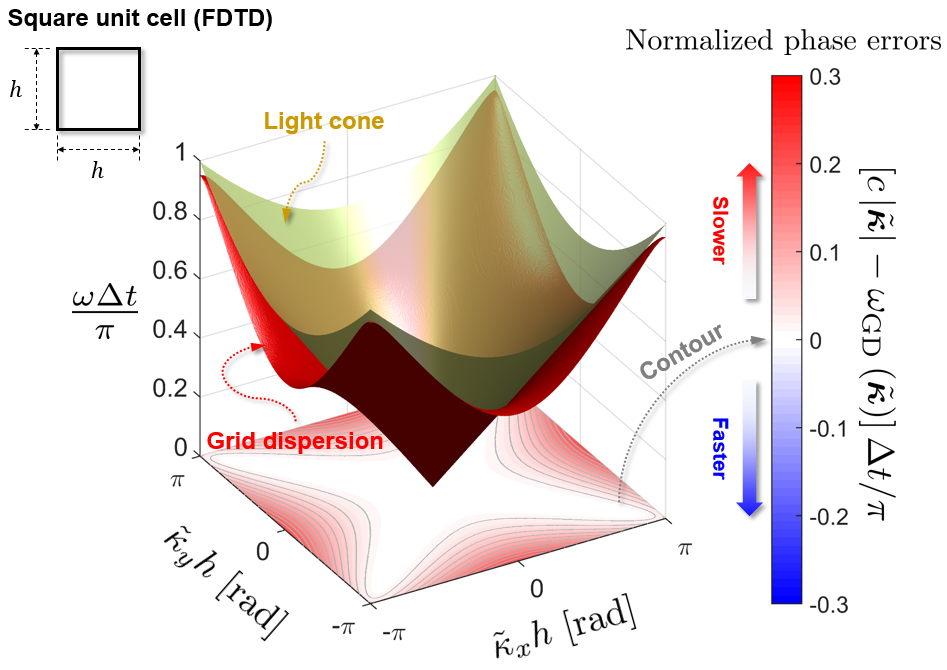}}
\subfloat[\label{fig:gd_fdtd_cfl_1_2d}]{\includegraphics[width=2in]{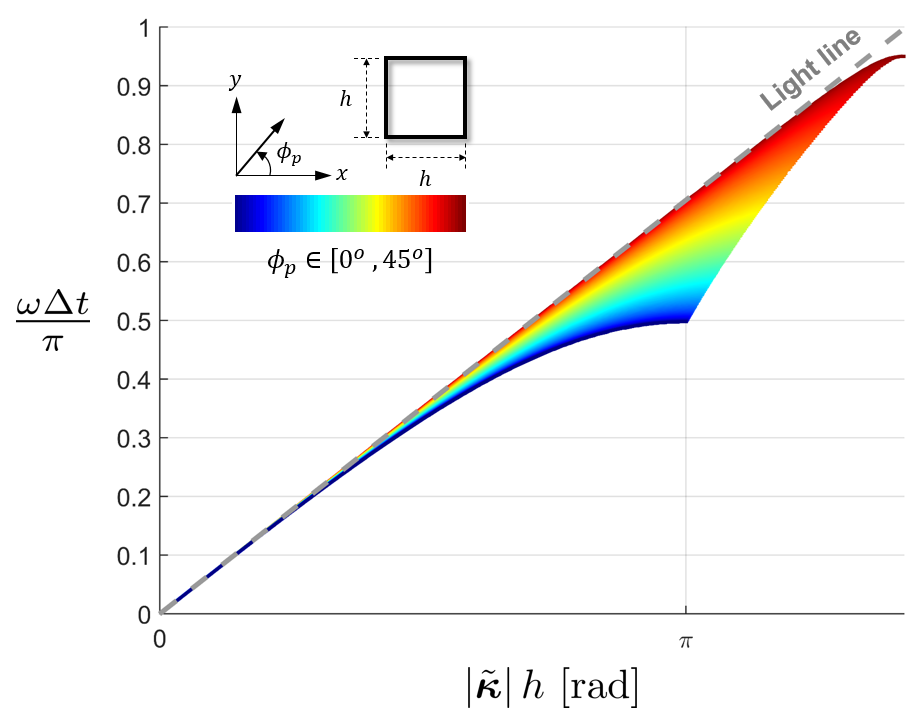}}
\caption{Numerical grid dispersion of the 2-D Yee's FDTD scheme on a structured mesh. (a) The red color surface represents the dispersion diagram of the normalized 
frequency $\omega\Delta t/\pi$ versus the normalized numerical wavenumber $\tilde{\boldsymbol{\kappa}}h$ in radians. The olive color surface represents the light cone. The contour levels at the bottom represent the normalized phase errors (with respect to the color bar). (b) Wavenumber magnitude versus frequency for different wave propagation angles with respect to the $x$ axis, $\phi_{p}\in\left[0^{o},45^{o}\right]$.}
\label{fig:gd_fdtd_cfl_1}
\end{figure}
The dispersion diagram $\left(\omega,\tilde{\boldsymbol{\kappa}}\right)$ can be plotted by solving (\ref{eqn:gd_fdtd}). Consider an example with $h=1$ m and $\Delta t=h/\left(\sqrt{2}c\right) \approx 2.35$ ns, which corresponds to the maximum time step allowed by the Courant-Friedrichs-Lewy (CFL) stability condition.
In Fig. \ref{fig:gd_fdtd_cfl_1_3d}, the numerical grid dispersion diagram $\omega_{\text{GD}}\left(\tilde{\boldsymbol{\kappa}}\right)$ is displayed by a red surface, and the light cone is shown in the olive green color. 
Contour levels at the bottom of the figure represent the normalized phase differences (errors) between the light cone and the numerical grid dispersion, $\left[c|\tilde{\boldsymbol{\kappa}}|-\omega_{\text{GD}}\left(\tilde{\boldsymbol{\kappa}}\right)\right]\Delta t/\pi$, indicative of how much faster or slower numerical waves propagate compared to the speed of light.
The normalized wave frequencies are plotted versus the normalized 
$\tilde{\boldsymbol{\kappa}}$
for various propagation directions
$\phi_{p}\in\left[0^{o},45^{o}\right]$ 
 along the grid\footnote{Here,
$\phi_{p}=0^{o}$ corresponds to a direction along the grid axis and 
$\phi_{p}=45^{o}$ corresponds to a direction along the grid diagonals. Due to the FDTD symmetry, the behavior in the $\phi_{p}\in\left[0^{o},45^{o}\right]$ repeats periodically along the other directions.} in Fig. \ref{fig:gd_fdtd_cfl_1_2d}. 
These results prove that the numerical phase velocity has an anisotropic behavior on the FDTD mesh and is always slower than the speed of light in vacuum.
Suppose that a cold plasma beam is relativistically drifting along the $x$-axis with a bulk beam velocity of $v_{\text{b}}=0.9 c$.
Its space-charge mode (or entropy mode) can be characterized in the dispersion diagram by a plane with inclination given by the beam velocity (beam plane).
NCR is emitted from the coupling region where the numerical grid dispersion surface and the beam plane intersect in the first Brillouin zone.
Furthermore, because of spatial and temporal aliasing effects, NCR can also be produced by resonances in other beam planes originated from higher-order Brillouin zones as well~\cite{godfrey2013numerical,xu2013numerical}, called {\it aliased beams}.  
The dispersion relation for space-charge modes {\color{black} of a cold electron} beam, including spatial and temporal aliasing effects, on a structured mesh is given by 
\begin{flalign}
\omega - \frac{2\pi}{\Delta t}v=v_{\text{b}} \left( \tilde{\kappa}_{x} - \frac{2\pi}{h}u \right)
\end{flalign}
where $u$ and $v$ are integers and the fundamental resonance mode has $u=v=0$ {\color{black}(see \ref{sec:beam_dispersion})}.
Since the spatial aliasing effect directly depends on the shape factors used for current deposition (scatter step) onto the FDTD grid, NCR can be mitigated to some extent  by increasing the spline order of the shape factors~\cite{ikeya2015stability}.

\begin{figure}
\centering
\subfloat[\label{fig:gd_fdtd_cfl_1_beam_3d_B_0_9}]{\includegraphics[width=2.2in]{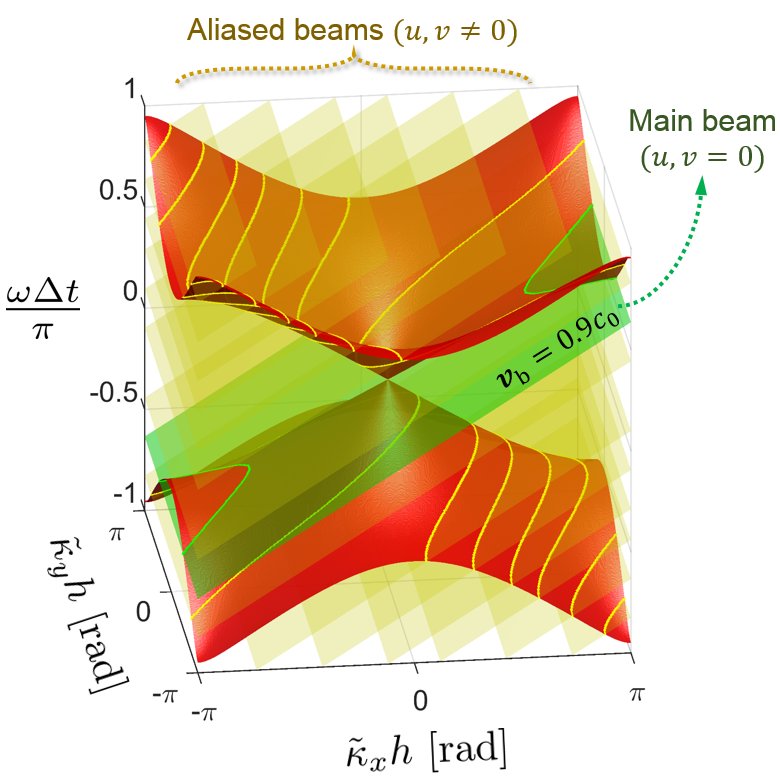}}
~
\subfloat[\label{fig:gd_fdtd_cfl_1_beam_2d_B_0_9}]{\includegraphics[width=2.5in]{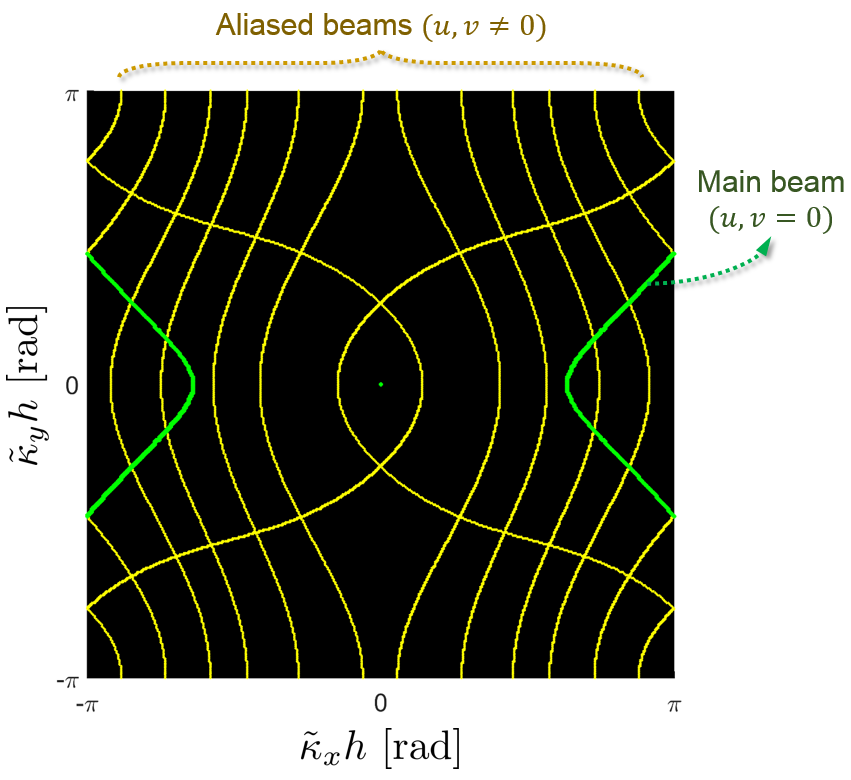}}	
\caption{Analytic NCR predictions on a structured FDTD grid for a bulk beam velocity $v_{\text{b}}=0.9c$. (a) 3-D numerical dispersion diagrams (in red) and beam planes (fundamental plane in green and aliased beams in transparent yellow). (b) Trajectories of NCR solutions projected onto the 2-D
 $\tilde{\boldsymbol{\kappa}}$-space.}
\label{fig:gd_fdtd_cfl_1_beam}
\end{figure}
Fig. \ref{fig:gd_fdtd_cfl_1_beam} illustrates the fundamental and aliased beam planes in the green and transparent yellow colors respectively, and the numerical dispersion in red within the first Brillouin zone, in which $\tilde{\boldsymbol{\kappa}}\in\left[-{\pi}/{h},{\pi}/{h}\right]\times\left[-{\pi}/{h},{\pi}/{h}\right]$ and $\omega\in\left[-{\pi}/{\Delta t},{\pi}/{\Delta t}\right]$.
As seen in Fig. \ref{fig:gd_fdtd_cfl_1_beam_3d_B_0_9}, the NCR, caused by the fundamental beam resonance, is present over a relatively narrow spectrum of wavenumbers due to the slower phase velocity of poorly-resolved waves for wavenumbers close to the grid cut-off.
On the other hand, NCR produced by aliased beam resonances  are spread out throughout the $\tilde{\boldsymbol{\kappa}}$-space and may occur regardless of whether numerical dispersion is corrected or not.
The loci for NCR solutions can be found by using root-find-solvers. These solutions are visualized more clearly in the
 $\tilde{\boldsymbol{\kappa}}$-space as depicted in Fig. \ref{fig:gd_fdtd_cfl_1_beam_2d_B_0_9} with $u,v \in \left\{-3,-2,...,2,3\right\}$. Again, the fundamental beam resonance is shown in green and aliased ones are shown in yellow.



\section{Numerical Cherenkov Radiation in finite-element-based EM-PIC Algorithms} \label{sec:FETD_Dispersion}
To analyze NCR on more general meshes, an EM-PIC algorithm~\cite{moon2015exact,na2016local,na2018relativistic} based on a mixed FETD field solver~\cite{kim2011parallel,he2006geometric} is adopted here for the reasons listed in the Introduction.
The present FETD field solver is based on an expansion of electromagnetic fields as a linear combination of  discrete differential forms (Whitney forms) 
defined over mesh elements and an explicit (leap-frog) discretization in time.
The most essential aspects of the field solver are outlined in \ref{sec:FETD_Derivation}.

\begin{figure}
	\centering
	\subfloat[\label{fig:sq_mesh_schm}]{\includegraphics[width=2in]{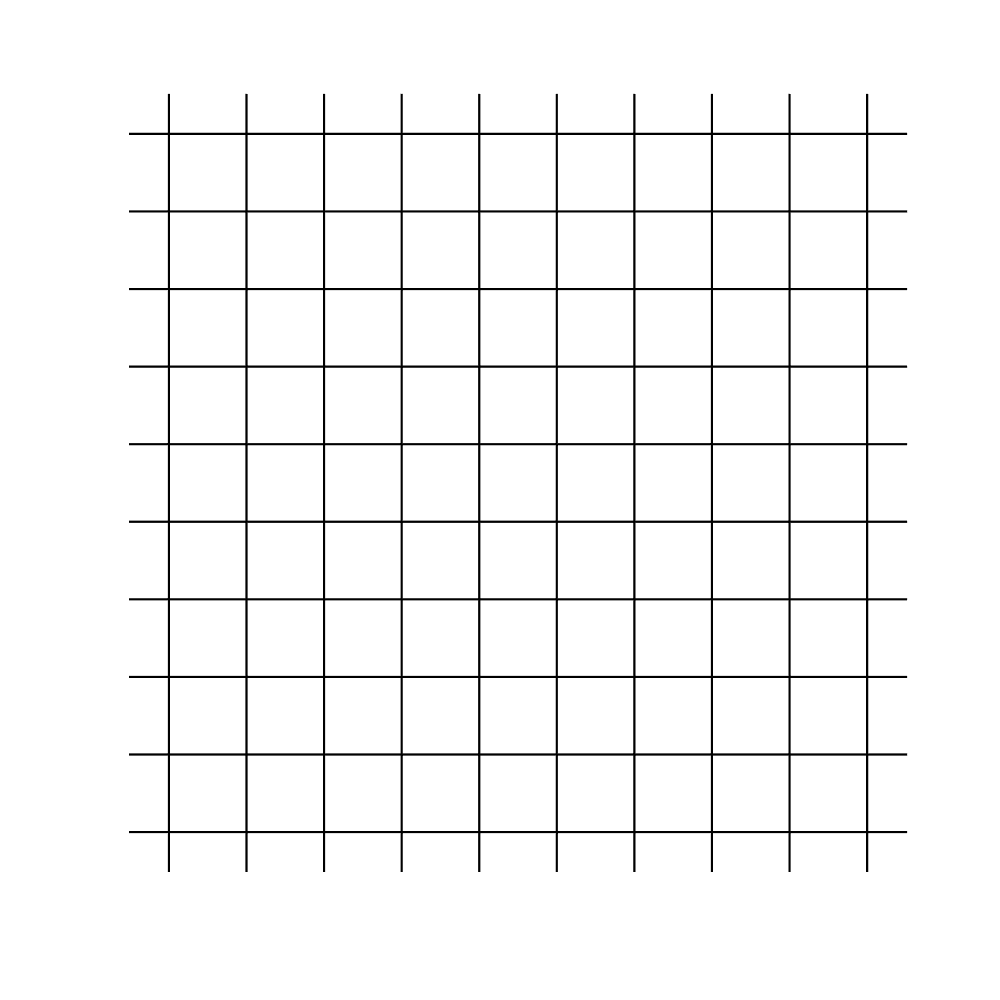}}
	\subfloat[\label{fig:rat_mesh_schm}]{\includegraphics[width=2in]{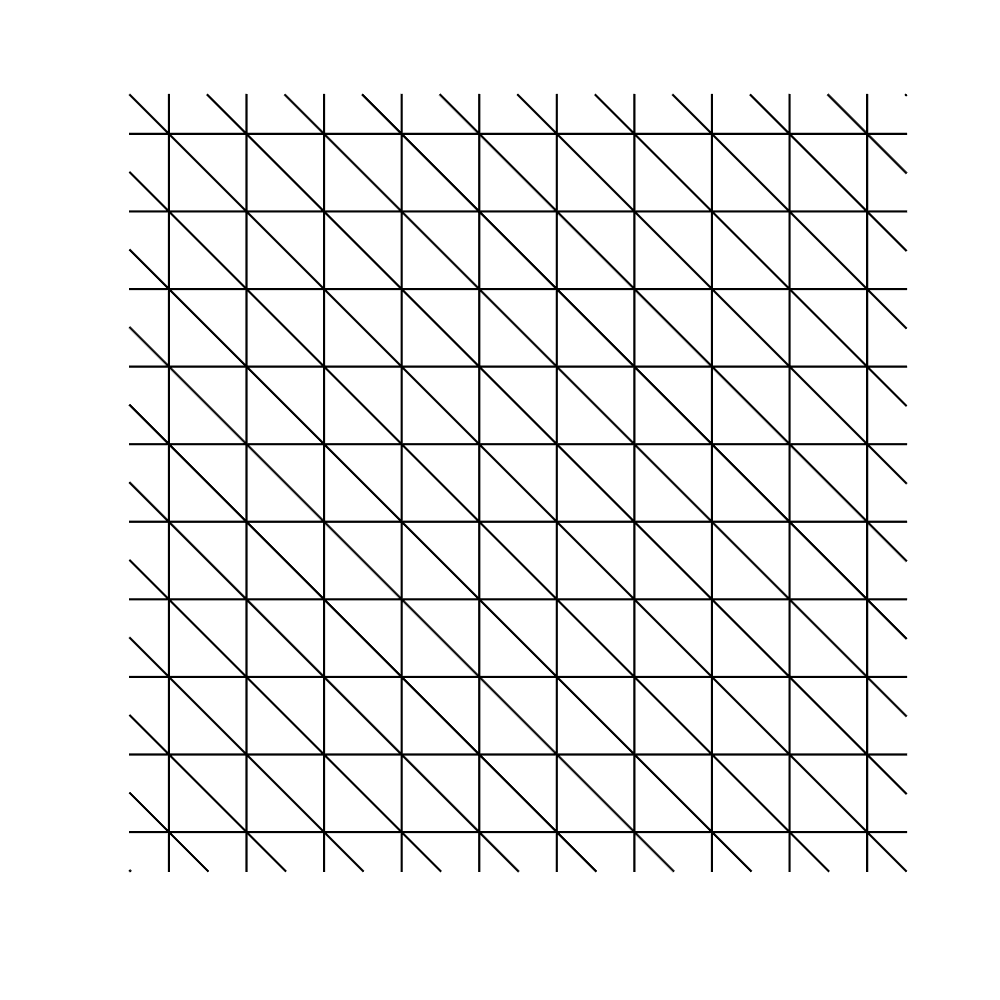}} \\
	\subfloat[\label{fig:isot_mesh_schm}]{\includegraphics[width=2in]{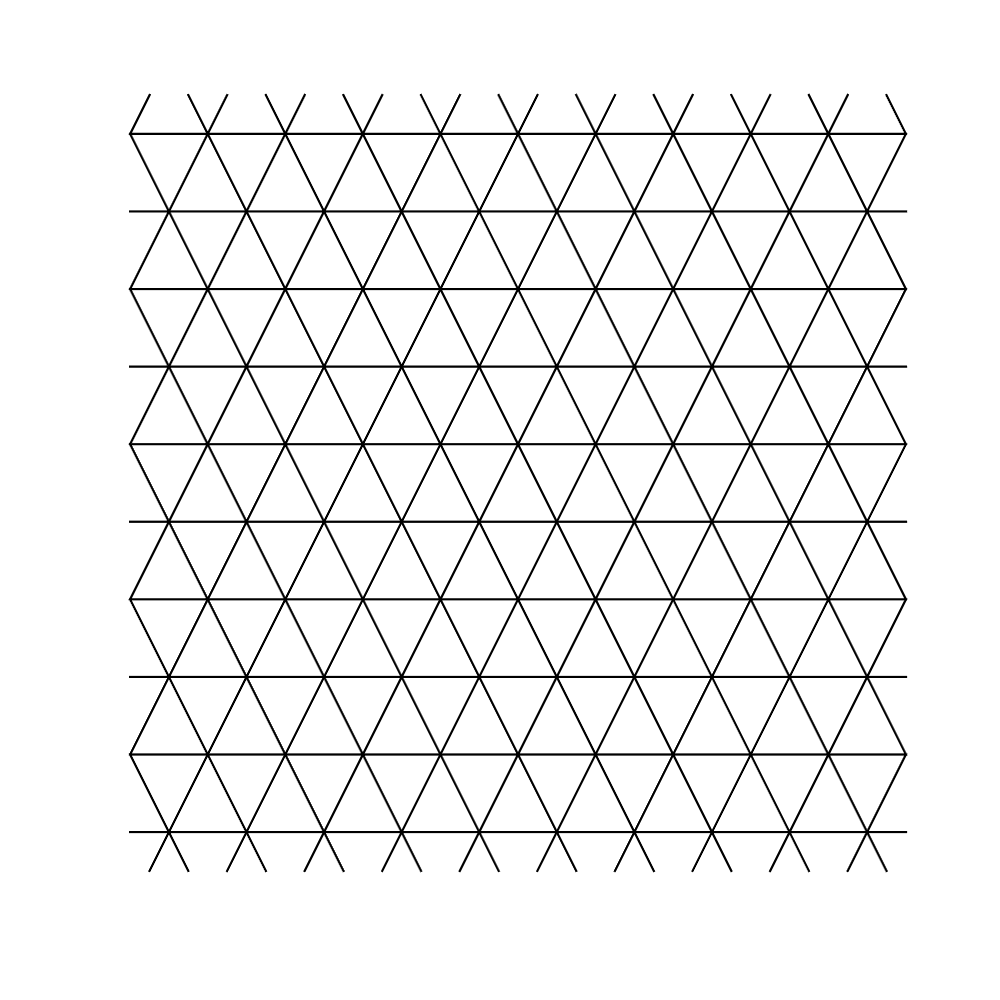}}
	\subfloat[\label{fig:higt_mesh_schm}]{\includegraphics[width=2in]{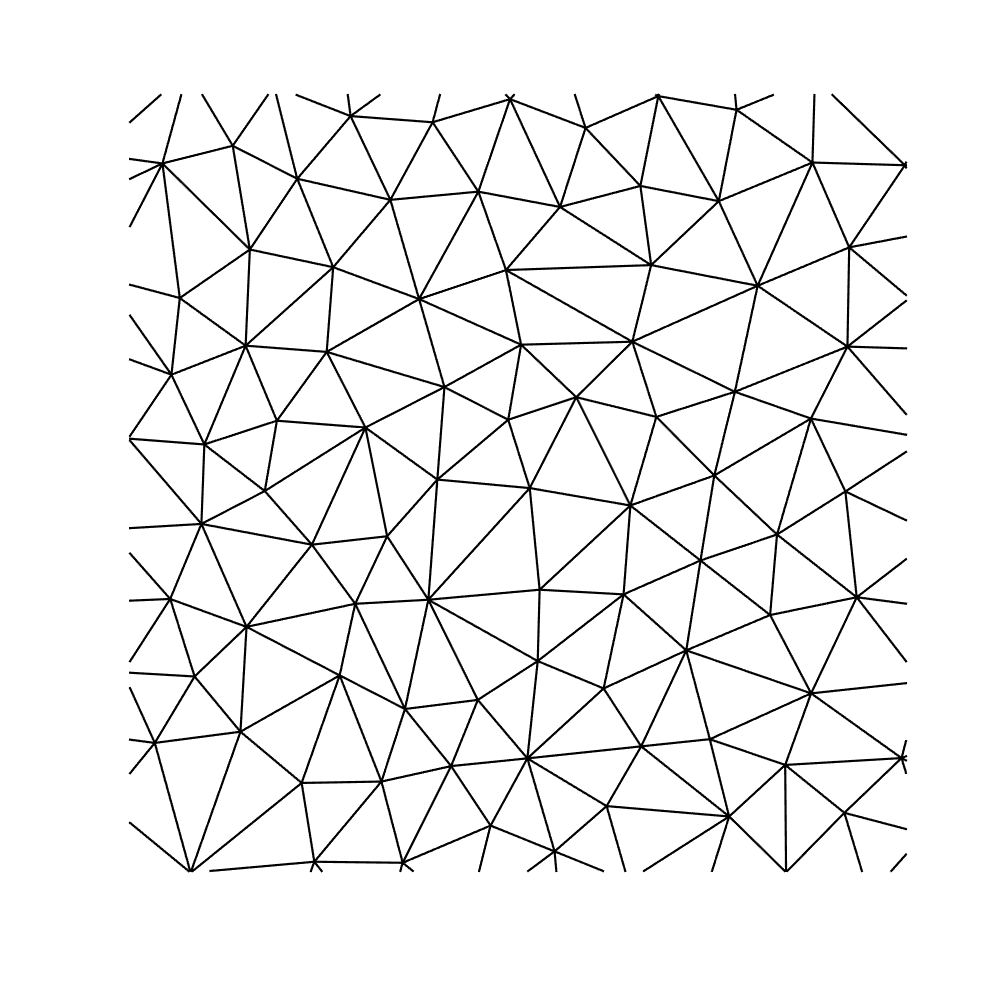}}
	\caption{Schematic illustration of the four types of mesh considered in this study. (a) Square regular (SQ) elements in both FDTD and FETD, (b) right-angle triangular (RAT) elements in FETD, (c) isosceles triangular (ISOT) elements in FETD, and (d) highly-irregular triangular (HIGT) elements in FETD.}
	\label{fig:mesh_schm}
\end{figure}

We first consider four periodic meshes, each with different element shapes and layouts as depicted in Fig. \ref{fig:mesh_schm}. They are denoted as square (SQ), right-angle triangular (RAT), isosceles triangular (ISOT), and highly-irregular triangular (HIGT). This section provides an {\it analytical} study of numerical dispersion and consequent NCR on SQ, RAT, and ISOT meshes. These three meshes have periodic arrangements of elements and hence are amenable to such analysis. Analytical predictions of this Section are compared against EM-PIC simulation results in the next Section. NCR effects in FETD-based EM-PIC simulations on HIGT meshes are also presented in the next Section.
The FETD-based EM-PIC algorithm is implemented on all four meshes, whereas the FDTD-based EM-PIC algorithm is implemented exclusively on the SQ mesh.

In order to analyze the numerical dispersion in the FETD algorithm implemented on SQ, RAT, and ISOT meshes, we consider a numerical plane wave expressed as $\mathbf{E}_{0}e^{j\left(\omega n \Delta t-\tilde{\boldsymbol{\kappa}}\cdot\mathbf{r}\right)}$. In the FETD algorithm, the discrete degrees of freedom (DoF) of the electric field are associated with the edges of the mesh.
Suppose that $\hat{t}_{j}$ is the tangential unit vector along the $j$\textsuperscript{th} edge of the mesh. 
The plane wave solutions can be projected onto the edges as $(\mathbf{E}_{0}\cdot \hat{t}_{j}) e^{j\left(\omega n \Delta t-\tilde{\boldsymbol{\kappa}}\cdot\mathbf{r}\right)}$ and the factor $\mathbf{E}_{0}\cdot \hat{t}_{j}$ can be taken as the DoF associated with the $j$\textsuperscript{th}edge. In addition, by using the superscript $n$ to represent the $n^{\text{th}}$ time step, we may denote the discrete DoFs as $e_{j}^{n}$.
For a plane wave propagating on a periodic mesh, it is possible to express the field value on an arbitrary edge using only the field values on a few number of so-called characteristic edges through multiplication of a spatial offset factor of the form $e^{-j\tilde{\boldsymbol{\kappa}}\cdot\overrightarrow{AA'}}$, where $\overrightarrow{AA'}$ is the relative position vector between the non-characteristic edge $A'$ and its corresponding characteristic edge $A$. Because of this, the number of DoFs can be restricted to the number of characteristic edges and the size of the matrices involved can be greatly reduced, see also~\cite{wu1997advantages}. Note that the SQ mesh has only two characteristic edges whereas the RAT and ISOT meshes have three characteristic edges. 

From \ref{sec:FETD_Derivation}, the full-discrete vector wave equation for the electric field{\color{black}, which is a second order explicit centered time discretization, can be written as \cite{kim2011parallel,moon2014trade} 
\begin{flalign}
\left[\star_{\epsilon}\right]\cdot\mathbf{e}^{n+1}
=
\left(2\left[\star_{\epsilon}\right]-\Delta t^2\mathbf{C}^{T}\cdot\left[\star_{\mu^{-1}}\right]\cdot\mathbf{C}\right)\cdot\mathbf{e}^{n}-\left[\star_{\epsilon}\right]\cdot\mathbf{e}^{n-1}.
\label{eq:st_disc_vec_wave_eq}
\end{flalign}
For a discrete plane wave with harmonic evolution of the form $e^{j\omega n \Delta t}$, it is clear that
 $\mathbf{e}^{n \pm 1}=\mathbf{e}^{n}e^{\pm j\omega \Delta t}$ so that (\ref{eq:st_disc_vec_wave_eq}) becomes
\begin{flalign}
\mathbf{X}\cdot\mathbf{e}^{n}
=\left\{2\left[\cos{\left(\omega\Delta t\right)}-1\right]\mathbf{M}+\Delta t^{2}\mathbf{S}\right\}\cdot\mathbf{e}^{n}
=\mathbf{0},
\label{eq:FETD_disp}
\end{flalign}
where $\mathbf{M}=\left[\star_{\epsilon}\right]$ (mass matrix) and $\mathbf{S}=\mathbf{C}^{T}\cdot\left[\star_{\mu^{-1}}\right]\cdot\mathbf{C}$ (stiffness matrix).
Non-trivial solutions can be obtained by solving $\det\left(\mathbf{X}\right)=0$
which determines the numerical dispersion relation $\left(\omega,\tilde{\boldsymbol{\kappa}}\right)$ on the mesh.
In what follows, NCR analysis is presented for SQ, RAT, and ISOT meshes. The numerical dispersion analysis is similar to \cite{wu1997advantages} except that time-discretization is also included.
\\
\subsection{SQ Mesh}\label{subsec:SQ_FETD_dispersion}
The SQ mesh has two characteristic edges, $y$- and $x$-directed. 
Let these two edges be denoted as $A$ and $B$, colored in red and blue, respectively, in Fig. \ref{fig:sq_uc_FETD}.
Local matrices for the three facets spanned by the support of the edge elements $A$ and $B$ are first computed. Then, the global mass and stiffness matrices can be constructed as a sum of three local matrices attributed to each facet.
The dashed red (for $y$-directed edges) and dashed blue (for $x$-directed edges) lines in Fig. \ref{fig:sq_uc_FETD} depict the relative position vectors between characteristic and non-characteristic edges.
\begin{figure}
\centering
\includegraphics[width=3in]{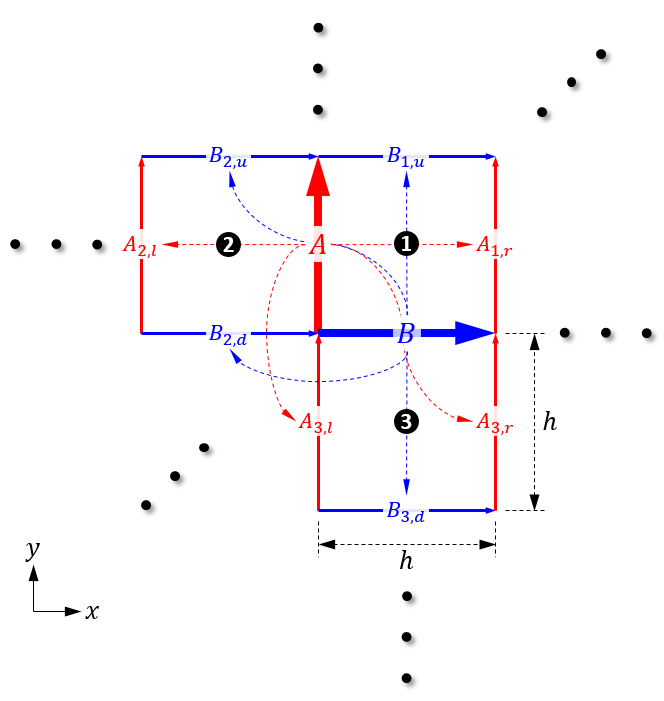}	
\caption{Schematic of SQ mesh. There are two characteristic edges ($A$ and $B$) directed along the $y$ and $x$ and colored in red and blue, respectively.}
\label{fig:sq_uc_FETD}
\end{figure}
As detailed in \ref{sec:FETD_Derivation}, in the mixed FETD algorithm, the electric field is represented as a linear combination of Whitney 1-forms associated 1:1 with mesh element edges and the magnetic flux density by a linear combination of Whitney 2-forms associated 1:1 with mesh element faces (note that Whitney 1- and 2-forms are also known as edge and face elements in the finite element literature).
The vector proxies of Whitney 1- and 2-forms for the $j$\textsuperscript{th} edge and $k$\textsuperscript{th} face are written as \cite{jin2015the}
\begin{flalign}
\mathbf{W}^{(1)}_{j}\left(\mathbf{r}\right)&=\hat{\alpha}\Pi\left(\frac{\alpha-\alpha_{j}^{(1)}}{h/2}\right)\Lambda\left(\frac{\beta-\beta_{j}^{(1)}}{h}\right),
\\
\mathbf{W}^{(2)}_{k}\left(\mathbf{r}\right)&=\frac{\hat{z}}{h^{2}}\Pi\left(\frac{x-x_{k}^{(2)}}{h/2}\right)\Pi\left(\frac{y-y_{k}^{(2)}}{h/2}\right),
\end{flalign}
where $j$ and $k$ are edge and face indices, $\alpha$ and $\beta$ stand for Cartesian coordinates $x$ and $y$ or vice-versa depending on the direction of the $j$\textsuperscript{th} edge, and the point $\left(x_{q}^{(p)},y_{q}^{(p)}\right)$ is the center of the $q$\textsuperscript{th} $p$-cell ($p=1$: edge, $p=2$: face). In addition, $\Pi\left(\cdot\right)$ and $\Lambda\left(\cdot\right)$ are scalar pulse and roof-top functions defined as
\begin{flalign}
\Pi\left(\zeta\right)&=
\left\{
\begin{array}{lr}
1, & \left|\zeta\right|\leq1\\
0, & \left|\zeta\right|>1
\end{array},
\right.
\\
\Lambda\left(\zeta\right)&=
\left\{
\begin{array}{lr}
1-\left|\zeta\right|, & \left|\zeta\right|\leq1\\
0, & \left|\zeta\right|>1
\end{array}
.
\right.
\end{flalign}
The global mass and stiffness matrices for DoFs on the characteristic edges, namely $\left[e_{A}^{n}, e_{B}^{n}\right]^{T}$, can then be written as
\begin{flalign}
\mathbf{M}&=\mathbf{M}_{1}+\mathbf{M}_{2}+\mathbf{M}_{3},
\\
\mathbf{S}&=\mathbf{S}_{1}+\mathbf{S}_{2}+\mathbf{S}_{3},
\end{flalign}
with
\begin{flalign}
\mathbf{M}_{1}&=\epsilon_{0}\Delta
\begin{bmatrix}
1/3 + e^{-j\tilde{\boldsymbol{\kappa}}\cdot\overrightarrow{AA}_{1,r}}/6 	& 0 \\
0 																		& 1/3 + e^{-j\tilde{\boldsymbol{\kappa}}\cdot\overrightarrow{BB}_{1,u}}/6
\end{bmatrix},
\nonumber \\
\mathbf{M}_{2}&=\epsilon_{0}\Delta
\begin{bmatrix}
1/3 + e^{-j\tilde{\boldsymbol{\kappa}}\cdot\overrightarrow{AA}_{2,l}}\Delta/6 	& 0 \\
0																			& 0
\end{bmatrix},
\nonumber \\
\mathbf{M}_{3}&=\epsilon_{0}\Delta
\begin{bmatrix}
0 & 0 \\
0 & 1/3 + e^{-j\tilde{\boldsymbol{\kappa}}\cdot\overrightarrow{BB}_{3,d}}/6
\end{bmatrix},
\\
\mathbf{S}_{1}&=\frac{1}{\mu_{0}\Delta}
\begin{bmatrix}
1-e^{-j\tilde{\boldsymbol{\kappa}}\cdot\overrightarrow{AA}_{1,r}}		& -1+e^{-j\tilde{\boldsymbol{\kappa}}\cdot\overrightarrow{BB}_{1,u}} \\
-1+e^{-j\tilde{\boldsymbol{\kappa}}\cdot\overrightarrow{AA}_{1,r}} 	& 1-e^{-j\tilde{\boldsymbol{\kappa}}\cdot\overrightarrow{BB}_{1,u}}
\end{bmatrix},
\nonumber \\
\mathbf{S}_{2}&=\frac{1}{\mu_{0}\Delta}
\begin{bmatrix}
1-e^{-j\tilde{\boldsymbol{\kappa}}\cdot\overrightarrow{AA}_{2,l}} 		& e^{-j\tilde{\boldsymbol{\kappa}}\cdot\overrightarrow{BB}_{2,d}}-e^{-j\tilde{\boldsymbol{\kappa}}\cdot\overrightarrow{BB}_{2,u}} \\
0 																	& 0
\end{bmatrix},
\nonumber \\
\mathbf{S}_{3}&=\frac{1}{\mu_{0}\Delta}
\begin{bmatrix}
0																															& 0 \\
e^{-j\tilde{\boldsymbol{\kappa}}\cdot\overrightarrow{AA}_{3,l}}-e^{-j\tilde{\boldsymbol{\kappa}}\cdot\overrightarrow{AA}_{3,r}},	& 1-e^{-j\tilde{\boldsymbol{\kappa}}\cdot\overrightarrow{BB}_{3,d}} 
\end{bmatrix},
\end{flalign}
where $\Delta$ is the area of the SQ mesh unit cell and the relative position vectors are given by $\overrightarrow{AA}_{1,r}=\left(h,0\right)$, $\overrightarrow{AA}_{2,l}=\left(-h,0\right)$, $\overrightarrow{AA}_{3,l}=\left(0,-h\right)$, $\overrightarrow{AA}_{3,r}=\left(h,-h\right)$, $\overrightarrow{BB}_{1,u}=\left(0,h\right)$, $\overrightarrow{BB}_{2,d}=\left(-h,0\right)$, $\overrightarrow{BB}_{2,u}=\left(-h,h\right)$, and $\overrightarrow{BB}_{3,d}=\left(0,-h\right)$.
{\color{black}
Again, the numerical grid dispersion can then be found by solving $\det\left(\mathbf{X}\right)=0$ and, at given $\tilde{\boldsymbol{\kappa}}$ on the first Brillouin zone, we numerically search zero-crossing points to make the residual of the determinant zero for the associated solution $\omega$.
}

\begin{figure}
\centering
\subfloat[\label{fig:gd_fetd_qdl_3d}]{\includegraphics[width=2.5in]{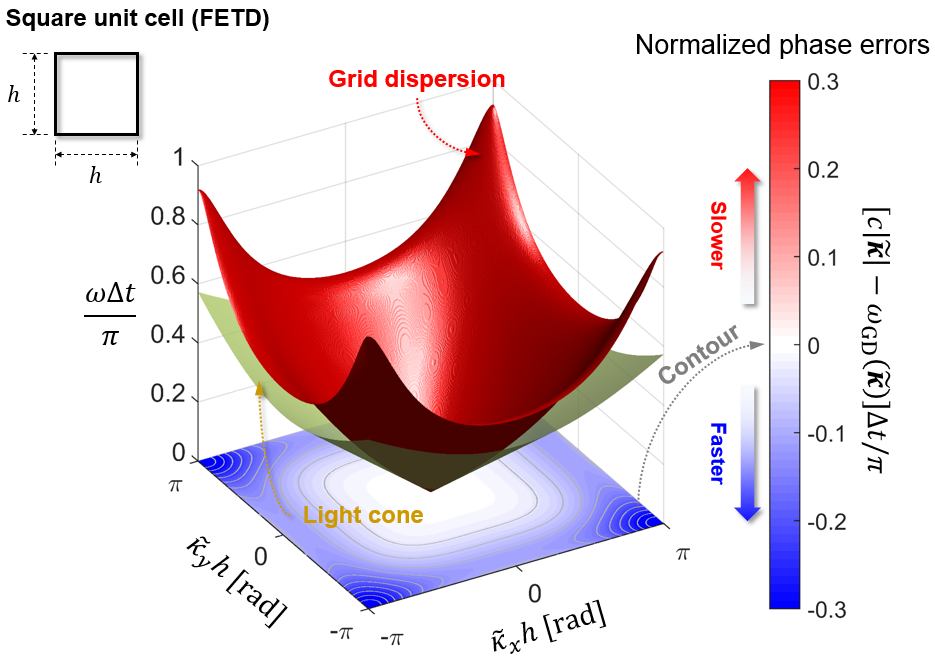}}
\subfloat[\label{fig:gd_fetd_qdl_2d}]{\includegraphics[width=2.1in]{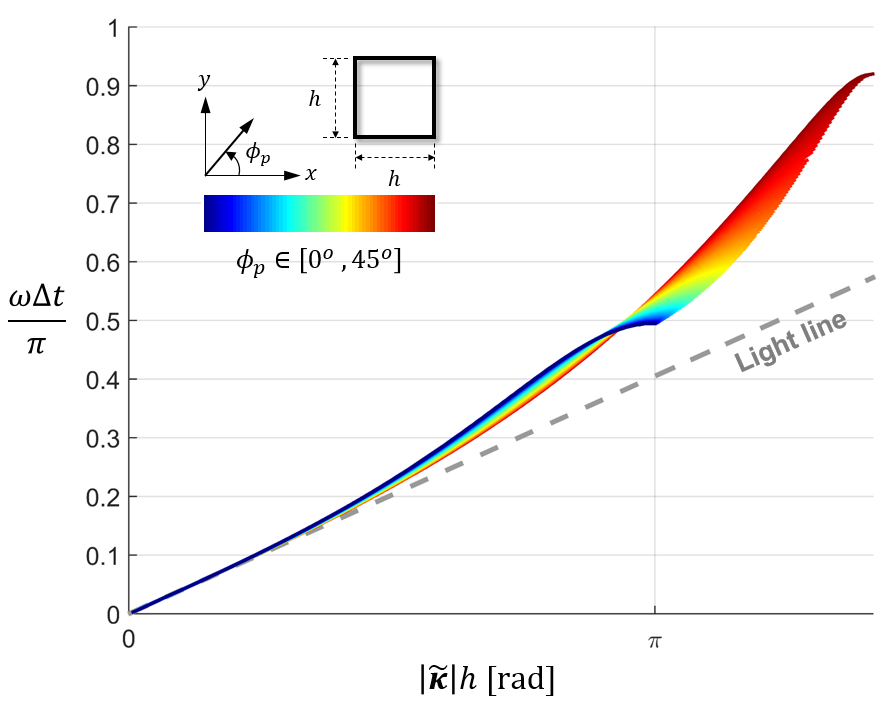}}
\caption{Numerical grid dispersion for the FETD scheme on the SQ mesh. (a) 
The red color surface represents the dispersion diagram of the normalized 
frequency $\omega\Delta t/\pi$ versus the normalized numerical wavenumber $\tilde{\boldsymbol{\kappa}}h$ in radians. The olive color surface represents the light cone. The contour levels at the bottom represent the normalized phase errors (with respect to the color bar). Note that the normalized phase error is always negative in this case because of a slightly faster-than-light numerical phase velocity.  (b) Projected dispersion curves for different wave propagation angles with respect to the $x$ axis $\phi_{p}\in\left[0^{o},45^{o}\right]$.}
\label{fig:gd_fetd_qdl_cfl_1}
\end{figure}
Fig.~\ref{fig:gd_fetd_qdl_cfl_1} illustrates the numerical grid dispersion of the FETD algorithm on the SQ mesh with $h=1$ [m] and $\Delta t=1.35$ [ns], which is the maximum time step for the stable field-update according to the CFL limit\footnote{The maximum time step in the FETD scheme can be obtained through an eigenvalue analysis on $\mathbf{M}^{-1}\cdot\mathbf{S}$ \cite{kim2011parallel}.}.
Unlike the FDTD case, it is clear from Fig. \ref{fig:gd_fetd_qdl_3d} that numerical plane waves propagate slightly faster than light in this case, regardless of the propagation direction, over the entire first Brillouin zone.
Fig.~\ref{fig:gd_fetd_qdl_2d} projects the numerical dispersion on the normalized wavenumber/frequency plane, with different propagation directions illustrated by different colors in the $\phi_{p}\in\left[0^{o},45^{o}\right]$ range.
As noted before, the standard FDTD algorithm can be recognized as a special version of the FETD on the SQ mesh in which low-order quadrature rules are employed in evaluating the mass (Hodge) matrices elements to yield diagonal matrices~\cite{lee2006note}. Conversely, the FETD algorithm on the SQ mesh can be regarded as a modified FDTD scheme in which non-diagonal coupling, present in the mass matrices, mimics an extended finite-difference stencil approximating spatial derivatives.
As discussed in~\cite{greenwood2004elimination}, the extended stencil makes numerical waves propagate faster than the speed of light. The latter effect can also be understood from the fact that the extended spatial stencil results in a stronger coupling of various degrees of freedom on the mesh.
Because of faster numerical wave speeds, the maximum time step for a stable update in the FETD algorithm on a SQ mesh is smaller than the FDTD limit on the same mesh by a factor of $0.58$, which agrees with \cite{greenwood2004elimination}.

Because the wave phase velocity of the FETD algorithm on the SQ mesh is larger than
$c$, it is expected that NCR will not arise from the wave resonance coupling to the fundamental beam plane. 
However, NCR is still expected in the solution domain due to the presence of spatially and temporally aliased beams, as illustrated in Fig. \ref{fig:gd_fetd_qdl_cfl_1_beam} with $u \in \left\{-5,-4,...,4,5\right\}$ and $v \in \left\{-3,-2,...,2,3\right\}$.
\begin{figure}
\centering
\subfloat[\label{fig:gd_fetd_qdl_cfl_1_b_0_9_beam_3d}]{\includegraphics[width=2.3in]{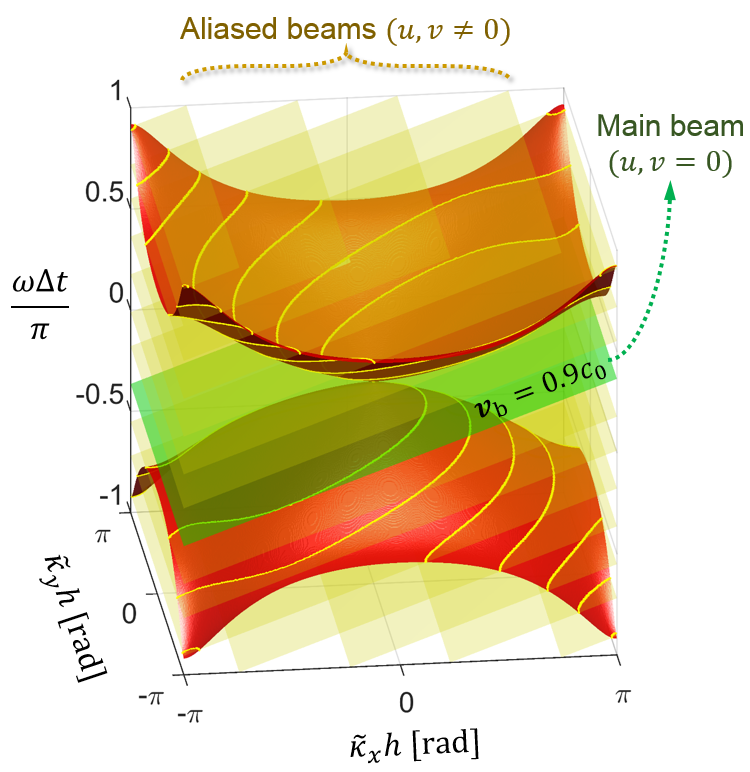}}
~
\subfloat[\label{fig:gd_fetd_qdl_cfl_1_b_0_9_beam_2d}]{\includegraphics[width=2.5in]{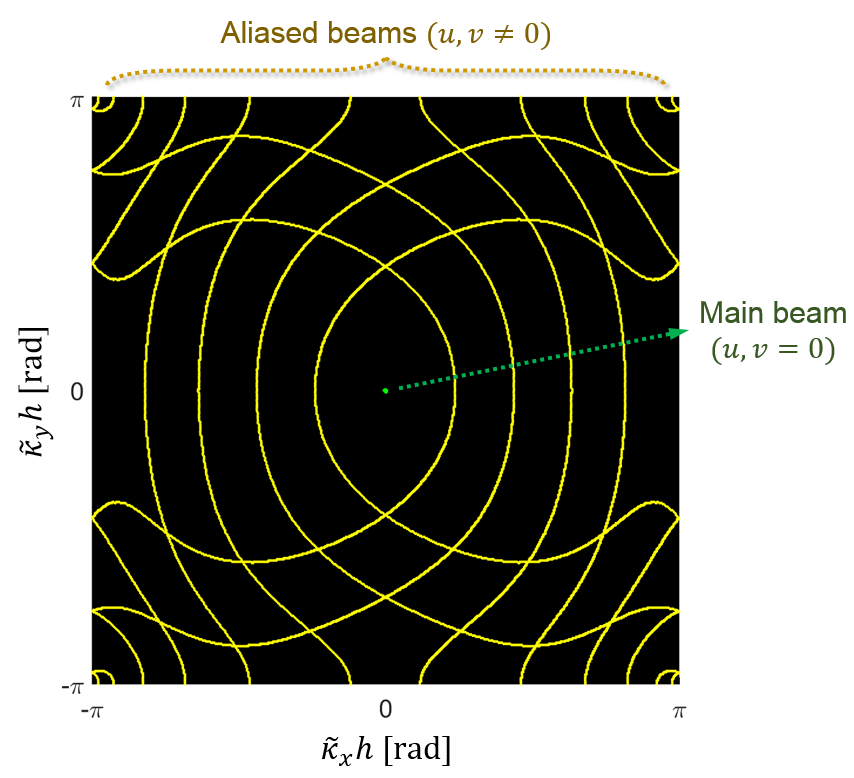}}	
\caption{Analytic prediction of NCR for the FETD algorithm on the SQ mesh when $v_{\text{b}}=0.9c$. (a) 3-D dispersion diagram. (b) NCR solution contours projected onto the first Brillouin zone in the  $\tilde{\boldsymbol{\kappa}}$-space.}
\label{fig:gd_fetd_qdl_cfl_1_beam}
\end{figure}
\\
\subsection{Triangular-element-based FE meshes}
On triangular-element-based FE meshes, we can again solve $\det\left(\mathbf{X}\right)=0$ to determine the numerical dispersion relation $\left(\omega,\tilde{\boldsymbol{\kappa}}\right)$ on the mesh.  However, the mass $\mathbf{M}$ and stiffness $\mathbf{S}$ matrices appearing on the expression for $\mathbf{X}$ should be modified. To derive these matrices,
we first recall the expression for the vector proxies of Whitney 1- and 2-forms on triangular meshes \cite{bossavit1988whitney,white2005mixed,rapetti2009whitney}
\begin{flalign}
\mathbf{W}^{(1)}_{j}\left(\mathbf{r}\right)&=\lambda_{{\left[1\right]}_{j}}\boldsymbol{\nabla}\lambda_{{\left[2\right]}_{j}}-\lambda_{{\left[2\right]}_{j}}\boldsymbol{\nabla}\lambda_{{\left[1\right]}_{j}}
\\
\mathbf{W}^{(2)}_{k}\left(\mathbf{r}\right)&=2\boldsymbol{\nabla}\lambda_{{\left\{1\right\}}_{k}}\times\boldsymbol{\nabla}\lambda_{{\left\{2\right\}}_{k}} (\text{on 2-dimensional meshes}).
\end{flalign}
at each $j$\textsuperscript{th} edge and $k$\textsuperscript{th} facet on the mesh, where  $\left[j\right]_{i}$ and  $\left\{j\right\}_{i}$ denote the $j$\textsuperscript{th} local node index for $i$\textsuperscript{th} edge and the $j$\textsuperscript{th} local node index for $i$\textsuperscript{th} facet, respectively. In addition, in what follows
$\left(j\right)_{i}$  denotes the $j$\textsuperscript{th} local edge index for $i$\textsuperscript{th} facet.
For an arbitrary point $\left(x,y\right)$ inside a $k$\textsuperscript{th} facet, the relationship between local nodal and barycentric coordinates is given by
\begin{flalign}
\begin{bmatrix}
x_{\left\{1\right\}_{k}} & x_{\left\{2\right\}_{k}} & x_{\left\{3\right\}_{k}}\\
y_{\left\{1\right\}_{k}} & y_{\left\{2\right\}_{k}} & y_{\left\{3\right\}_{k}}\\
1 & 1 & 1\\
\end{bmatrix}
\cdot
\begin{bmatrix}
\lambda_{\left\{1\right\}_{k}}\\
\lambda_{\left\{2\right\}_{k}}\\
\lambda_{\left\{3\right\}_{k}}\\
\end{bmatrix}
=
\begin{bmatrix}
x\\
y\\
1\\
\end{bmatrix}
.
\end{flalign}

\begin{figure}
\centering
\includegraphics[width=3.25in]{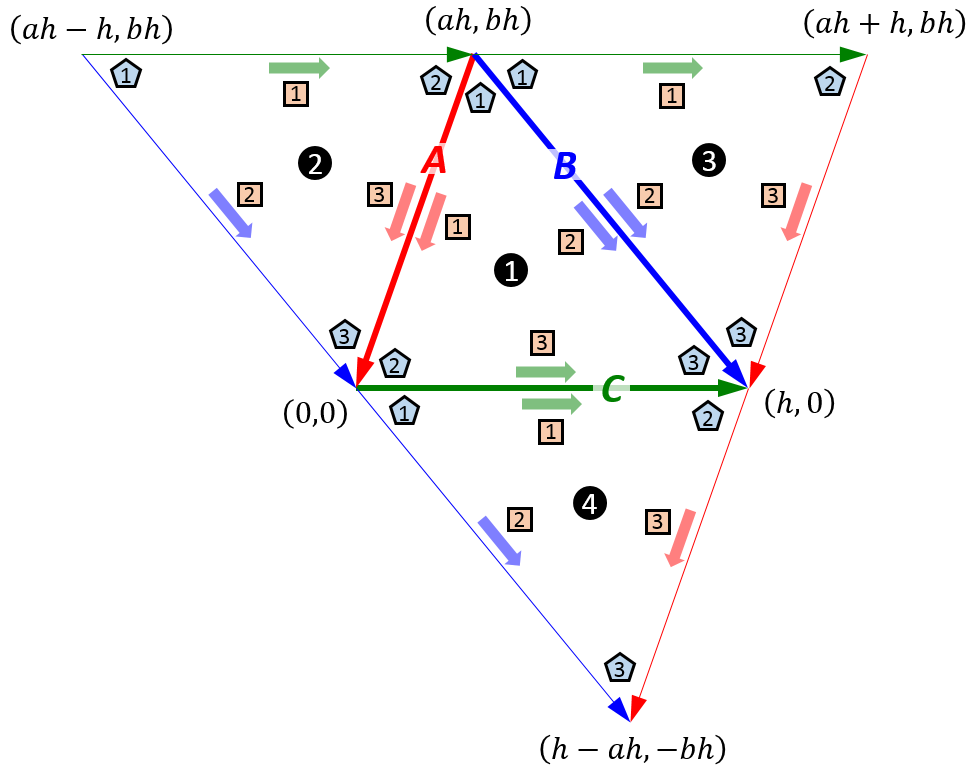}	
\caption{A periodically-arranged triangular grid. It has three characteristic edges denoted by $A$, $B$, and $C$. Labels inside circles denote global facet indexes and labels inside rectangles and pentagons denote local edge and node indexes, respectively.}
\label{fig:tr_uc_FETD}
\end{figure}
Consider a periodically-arranged triangular grid, as shown in Fig. \ref{fig:tr_uc_FETD}.
The local mass matrix for the $k$\textsuperscript{th} facet can be obtained by \cite{strang2008analysis,jin2015the}
\begin{flalign}
\mathbf{M}^{\text{loc}}_{k}=
\begin{bmatrix}
M^{\text{loc}}_{\left(1,1\right)_{k}} & M^{\text{loc}}_{\left(1,2\right)_{k}} & M^{\text{loc}}_{\left(1,3\right)_{k}} \\
M^{\text{loc}}_{\left(2,1\right)_{k}} & M^{\text{loc}}_{\left(2,2\right)_{k}} & M^{\text{loc}}_{\left(2,3\right)_{k}} \\
M^{\text{loc}}_{\left(3,1\right)_{k}} & M^{\text{loc}}_{\left(3,2\right)_{k}} & M^{\text{loc}}_{\left(3,3\right)_{k}}
\end{bmatrix},
\end{flalign}
where
\begin{flalign}
M^{\text{loc}}_{\left(i,j\right)_{k}}=2A_{k}
\begin{bmatrix}
\alpha_{i,j} & \beta_{i,j} &\gamma_{i,j}
\end{bmatrix}
\cdot
\begin{bmatrix}
\boldsymbol{\nabla}\lambda_{\left\{1\right\}_{k}} \cdot \boldsymbol{\nabla}\lambda_{\left\{1\right\}_{k}}\\
\boldsymbol{\nabla}\lambda_{\left\{1\right\}_{k}} \cdot \boldsymbol{\nabla}\lambda_{\left\{2\right\}_{k}}\\
\boldsymbol{\nabla}\lambda_{\left\{2\right\}_{k}} \cdot \boldsymbol{\nabla}\lambda_{\left\{2\right\}_{k}}
\end{bmatrix},
\end{flalign}
\begin{flalign}
\boldsymbol{\alpha}=\frac{1}{12}
\begin{bmatrix}
1 & 1 & 1 \\
1 & 3 & 1 \\
1 & 3 & 1
\end{bmatrix}
,
\boldsymbol{\beta}=\frac{1}{12}
\begin{bmatrix}
-1 & -1 & 1 \\
-1 & 3 & 3 \\
1 & 3 & 3
\end{bmatrix}
,
\boldsymbol{\gamma}=\frac{1}{12}
\begin{bmatrix}
1 & -1 & -1 \\
-1 & 1 & 1 \\
-1 & 1 & 3
\end{bmatrix}
.
\end{flalign}
and $A_{k}$ is the area of the $k$\textsuperscript{th} facet. Each element index in the expression of the local mass matrix above also denotes the local edge index of the $k$\textsuperscript{th} facet. 
The local stiffness matrix $\mathbf{S}^{\text{loc}}_{k}=\left(\mathbf{C}^{\text{loc}}_{k}\right)^{T}\cdot\left[\star_{\mu^{-1}}\right]^{\text{loc}}_{k}\cdot\mathbf{C}^{\text{loc}}_{k}$ is given by
\begin{flalign}
\mathbf{S}^{\text{loc}}_{k^{\text{th}}}=
\begin{bmatrix}
S^{\text{loc}}_{\left(1,1\right)_{k}} & S^{\text{loc}}_{\left(1,2\right)_{k}} & S^{\text{loc}}_{\left(1,3\right)_{k}} \\
S^{\text{loc}}_{\left(2,1\right)_{k}} & S^{\text{loc}}_{\left(2,2\right)_{k}} & S^{\text{loc}}_{\left(2,3\right)_{k}} \\
S^{\text{loc}}_{\left(3,1\right)_{k}} & S^{\text{loc}}_{\left(3,2\right)_{k}} & S^{\text{loc}}_{\left(3,3\right)_{k}}
\end{bmatrix},
\end{flalign}
where $S^{\text{loc}}_{\left(i,j\right)_{k}}={\left(-1\right)^{i+j}}/{A_{k}}$.
We define a function $F\left(\left\{j\right\}_{k},k\right)$ that yields one of the characteristic edges of the $k$\textsuperscript{th} facet according to the local edge index $\left\{j\right\}_{k}$, i.e., $F\left(\left\{j\right\}_{k},k\right)=g_{k}(=A$, $B$, or $C$). Likewise, the inverse function $F^{-1}=H$ yields the $j$\textsuperscript{th} local edge index of the $k$\textsuperscript{th} facet, $\left\{j\right\}_{k}$, i.e. $H\left(F\left(\left\{j\right\}_{k},k\right)\right)=\left\{j\right\}_{k}$.

Projecting plane wave solutions into DoFs on edges, the relationship between the DoFs for similar edges in different facets 
$k$\textsuperscript{th} and $K$\textsuperscript{th} can be written as
\begin{flalign}
e_{g_{K}}=e_{g_{k}}e^{-j\tilde{\boldsymbol{\kappa}}\cdot\overrightarrow{g_{k}g_{K}}}=e_{F\left(\left\{j\right\}_{k},k\right)}e^{-j\varphi_{[g_{k}\rightarrow g_{K}]}}
\end{flalign}
where $\overrightarrow{g_{k}g_{K}}$ is the relative position vector from the center of the edge $g_{k}$ to $g_{K}$ and $\varphi_{[g_{k}\rightarrow g_{K}]}$ is the corresponding phase delay.
For $\mathbf{e}=\left[e_{A},e_{B},e_{C}\right]^{T}$, the global mass and stiffness matrices can be assembled by
\begin{flalign}
M_{G_{1},g_{1}}&=\sum_{p=1}^{4}
\left\{
\begin{array}{lr}
M^{\text{loc}}_{\left(H\left(G_{p}\right),H\left(g_{p}\right)\right)_{p}} & \text{, for } G_{p}=g_{p}\\
M^{\text{loc}}_{\left(H\left(G_{p}\right),H\left(g_{p}\right)\right)_{p}}e^{-j\varphi_{[g_{1}\rightarrow g_{p}]}} & \text{, for } G_{p}\neq g_{p}\\
\end{array}
\right.
,
\\
S_{G_{1},g_{1}}&=\sum_{p=1}^{4}
\left\{
\begin{array}{lr}
S^{\text{loc}}_{\left(H\left(G_{p}\right),H\left(g_{p}\right)\right)_{p}} & \text{, for } G_{p}=g_{p}\\
S^{\text{loc}}_{\left(H\left(G_{p}\right),H\left(g_{p}\right)\right)_{p}}e^{-j\varphi_{[g_{1}\rightarrow g_{p}]}} & \text{, for } G_{p}\neq g_{p}\\
\end{array}
\right.
,
\end{flalign}
where $G_{p}$ and $g_{p}$ are outputs of the $F$ function for local edges in $p$ facets (see Fig. \ref{fig:tr_uc_FETD}). 
As noted before, the numerical grid dispersion of the FETD schemes on periodically-arranged triangular grids can then be obtained by substituting the above mass and stiffness matrices into (\ref{eq:FETD_disp}) and solving the characteristic equation $\det\left(\mathbf{X}\right)=0$. We will examine next triangular meshes composed of RAT and ISOT elements.
\\
\subsubsection{Right-angle triangular-element (RAT) mesh}
The unit cell of the RAT mesh corresponds to setting $a=0$ and $b=1$ (see Fig. \ref{fig:tr_uc_FETD}).
\begin{figure}
\centering
\subfloat[\label{fig:gd_fetd_rat_3d}]{\includegraphics[width=2.7in]{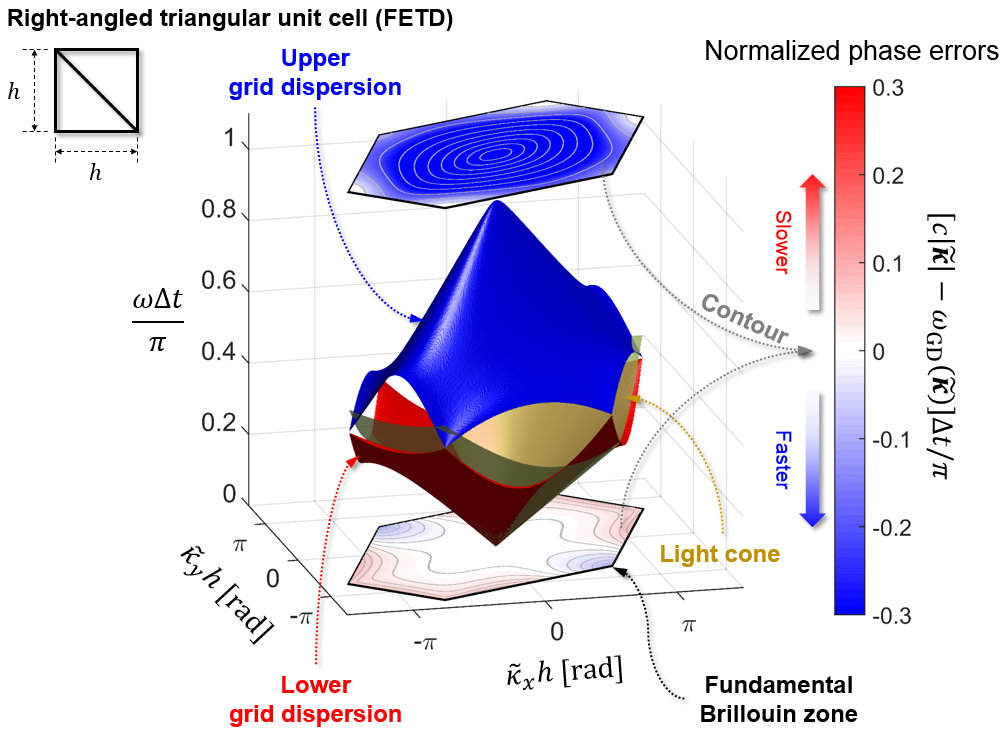}}
\subfloat[\label{fig:gd_fetd_rat_2d}]{\includegraphics[width=2.1in]{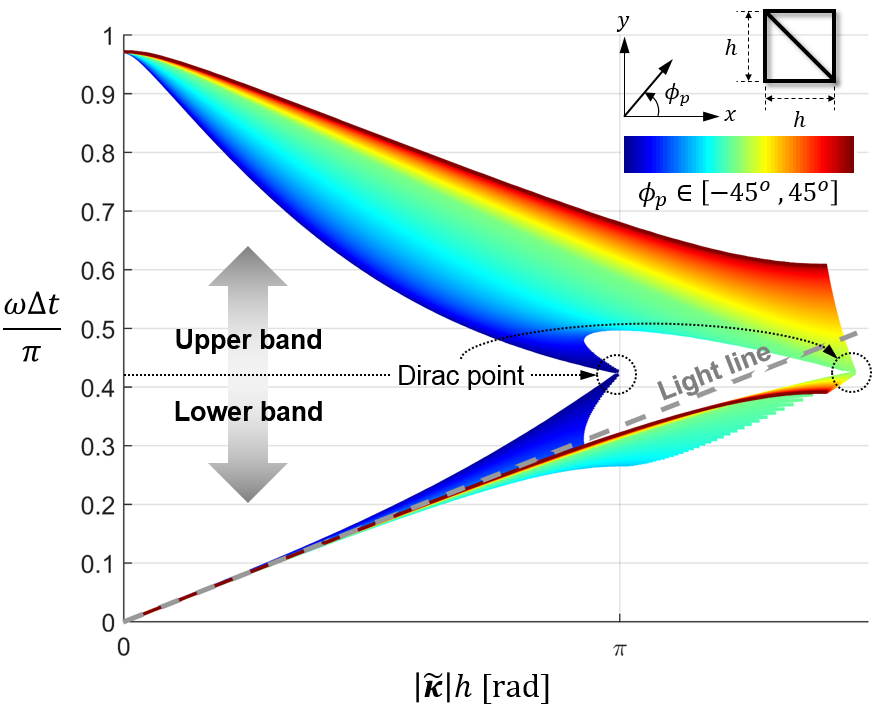}}
\caption{Numerical grid dispersion for the FETD scheme on the RAT mesh with the CFL number equal to one. Unlike the FDTD or FETD-SQ cases, this 
diagram exhibits an additional (upper) dispersion band. (a) 
The red (lower band) and blue (upper band) color surfaces represent the dispersion diagram of the normalized 
frequency $\omega\Delta t/\pi$ versus the normalized numerical wavenumber $\tilde{\boldsymbol{\kappa}}h$ in radians. The olive color surface represents the light cone. The contour levels at the bottom represent the normalized phase errors (with respect to the color bar).  (b) Projected dispersion curves for different wave propagation angles with respect to the $x$ axis $\phi_{p}\in\left[-45^{o},45^{o}\right]$.}
\label{fig:gd_fetd_rat_cfl_1}
\end{figure}
The numerical dispersion diagram of the FETD-RAT scheme over the first spatial Brillouin zone is illustrated in Fig. \ref{fig:gd_fetd_rat_cfl_1}.
The first spatial Brillouin zone of the RAT mesh in the $\tilde{\boldsymbol{\kappa}}$-space is shaped as an inclined hexagon.
{\color{black}
Unlike the SQ meshes with FD- and FETD schemes, two $\omega$ roots of the dispersion relations are found for each $\tilde{\boldsymbol{\kappa}}$ in this mesh.}
As depicted in Fig. \ref{fig:gd_fetd_rat_3d}, red and blue surfaces correspond to the lower and upper grid dispersion of the mesh. The normalized phase errors are shown by means of contour maps at the bottom and top planes along the (vertical) frequency axis.
Fig. \ref{fig:gd_fetd_rat_2d} shows the numerical dispersion curves for different propagation directions $\phi_{p}\in\left[-45^{o},45^{o}\right]$.
Dirac points denote the points where the lower and upper dispersion bands meet\footnote{This nomenclature is borrowed from solid state physics, where it is used to describe attachment of valence and conduction energy bands.}. 
It is observed that the numerical dispersion in the lower band is highly anisotropic.
Although in the upper band wave propagation is faster than light due to its inverse-like shape compared to the lower band, NCR may still be produced by intersection with aliased beams.
The existence of an upper dispersion band on meshes based on RAT elements can be understood from the fact that
the normal component of the (vector proxy of) Whitney 1-forms used to expand the electric field on the mesh exhibit discontinuities at the edges of triangular meshes. 
Fig. \ref{fig:wh_1_norm_discont} illustrates the normal discontinuity of Whitney 1-forms. 
This is in contrast to meshes based on SQ elements where
Whitney 1-forms exhibit both tangential and normal continuity (due to zero normal components). Strictly speaking, Whitney 1-forms on triangular grids are only tangentially continuous \cite{bossavit1988whitney,rapetti2009whitney}.
Indeed, the numerical grid dispersion behavior on meshes with periodically-arranged triangular elements is reminiscent of that of photonic band gap structures in which the discontinuity of the normal field component is caused by periodic material interfaces.

\begin{figure}
\centering
\subfloat[\label{fig:wh_1_vec}]{\includegraphics[width=1.5in]{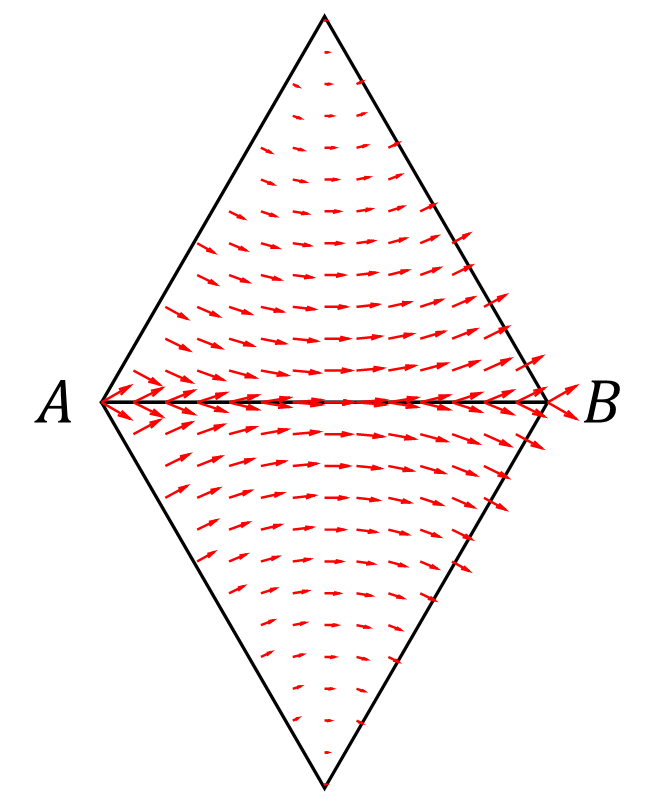}}
~
\subfloat[\label{fig:wh_1_vec_t}]{\includegraphics[width=1.5in]{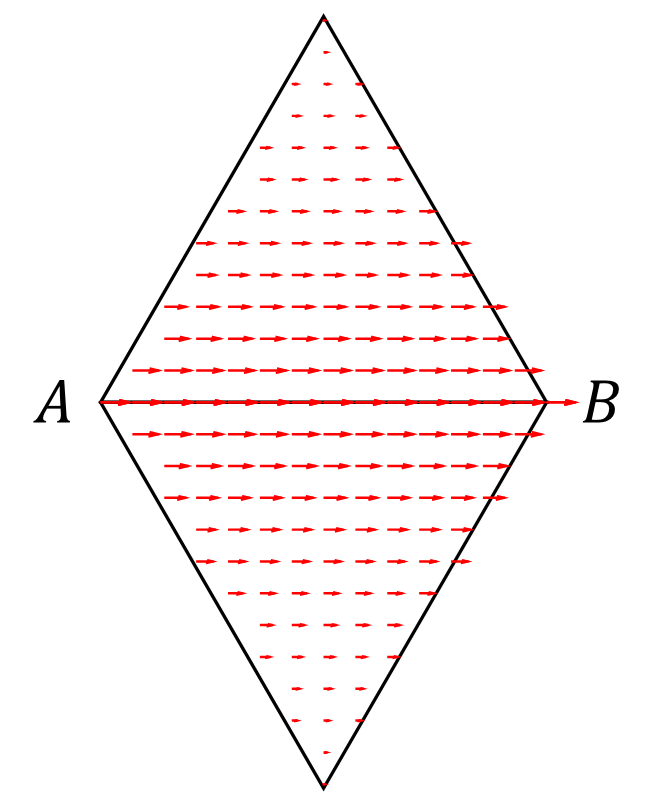}}	
~
\subfloat[\label{fig:wh_1_vec_n}]{\includegraphics[width=1.5in]{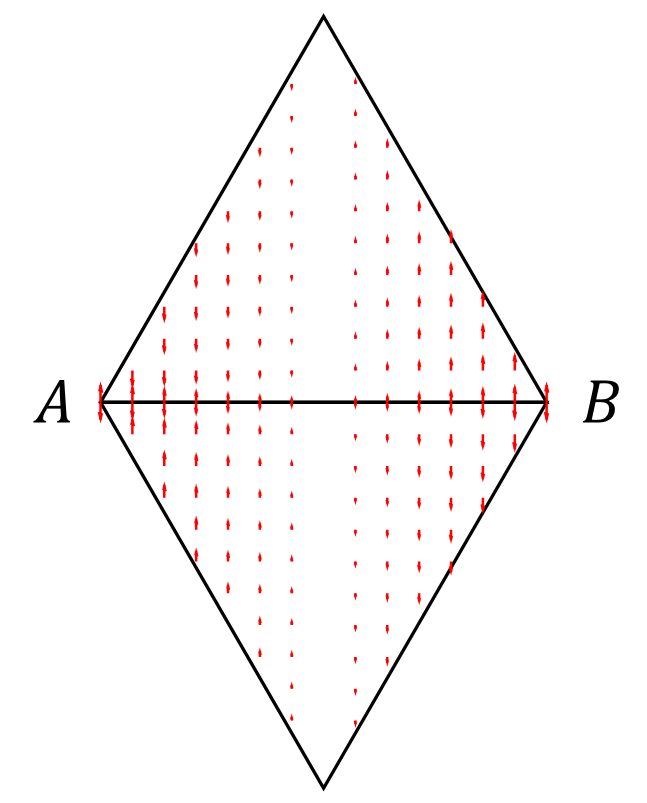}}	
\caption{(a) The vector proxy of a Whitney 1-form associated with the edge $\protect\overrightarrow{AB}$ on a triangular mesh. (b) Tangential component along edge. (c) Normal component to the edge direction. }
\label{fig:wh_1_norm_discont}
\end{figure}

\begin{figure}
\centering
\subfloat[\label{fig:gd_fetd_rat_cfl_1_b_0_9_beam_3d}]{\includegraphics[width=2.55in]{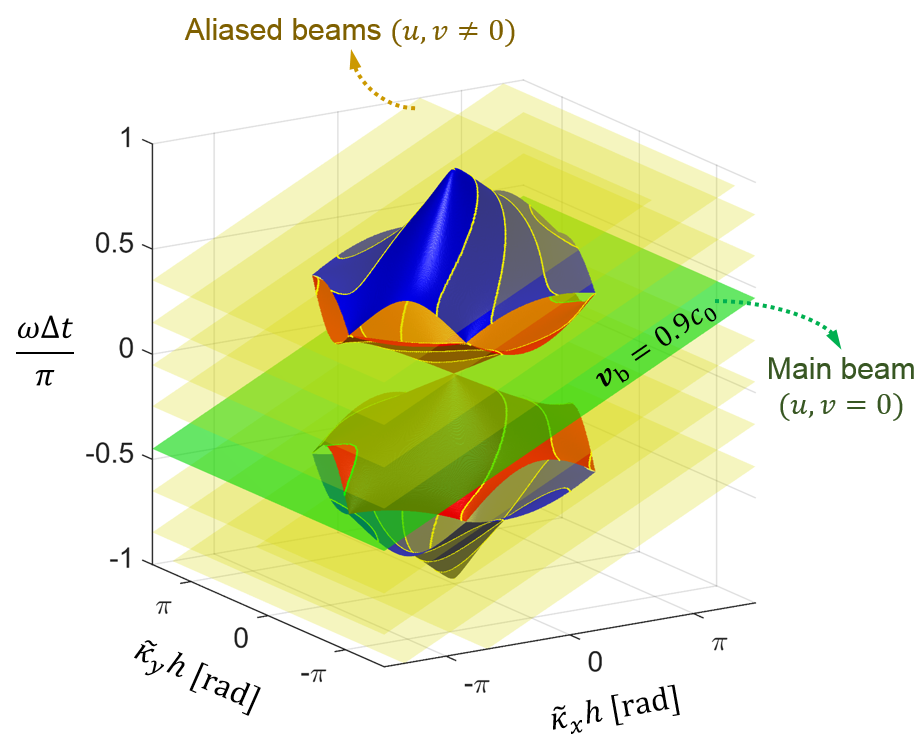}}
~~
\subfloat[\label{fig:gd_fetd_rat_cfl_1_b_0_9_beam_2d}]{\includegraphics[width=2.1in]{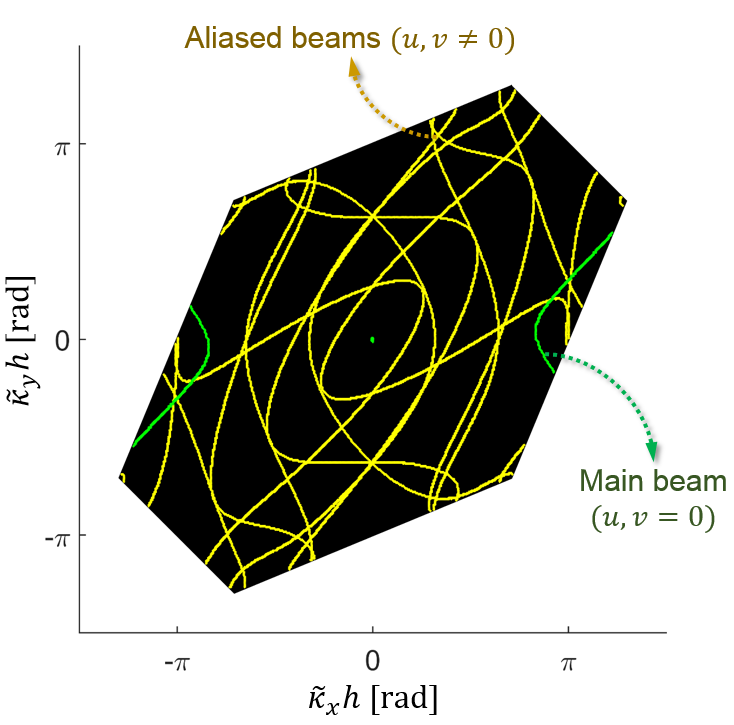}}	
\caption{Analytic prediction of NCR for the FETD-based EM-PIC scheme on the RAT mesh assuming a plasma beam with bulk velocity $v_{\text{b}}=0.9c$. 
(a) Dispersion diagram. (b) NCR solution contours projected onto the first Brillouin zone in the 
 $\tilde{\boldsymbol{\kappa}}$-space.}
\label{fig:gd_fetd_rat_cfl_1_beam}
\end{figure}
Fig. \ref{fig:gd_fetd_rat_cfl_1_beam} shows the analytic prediction of NCR for the FETD-based EM-PIC scheme on the RAT mesh  assuming a plasma beam with bulk velocity $v_{\text{b}}=0.9c$. 
Fig. \ref{fig:gd_fetd_rat_cfl_1_b_0_9_beam_3d} depicts the numerical dispersion diagram over the first Brillouin zone superimposed to the fundamental and aliased beams. 
Set of NCR solutions in the $\tilde{\boldsymbol{\kappa}}$-space
  are shown in Fig. \ref{fig:gd_fetd_rat_cfl_1_b_0_9_beam_2d} with $u \in \left\{-5,-4,...,4,5\right\}$ and $v \in \left\{-3,-2,...,2,3\right\}$.
\\
\subsubsection{Isosceles triangular-element (ISOT) mesh}
The unit cell for the ISOT mesh corresponds to $a=0.5$ and $b=1$.
\begin{figure}
\centering
\subfloat[\label{fig:gd_fetd_isom_3d}]{\includegraphics[width=2.65in]{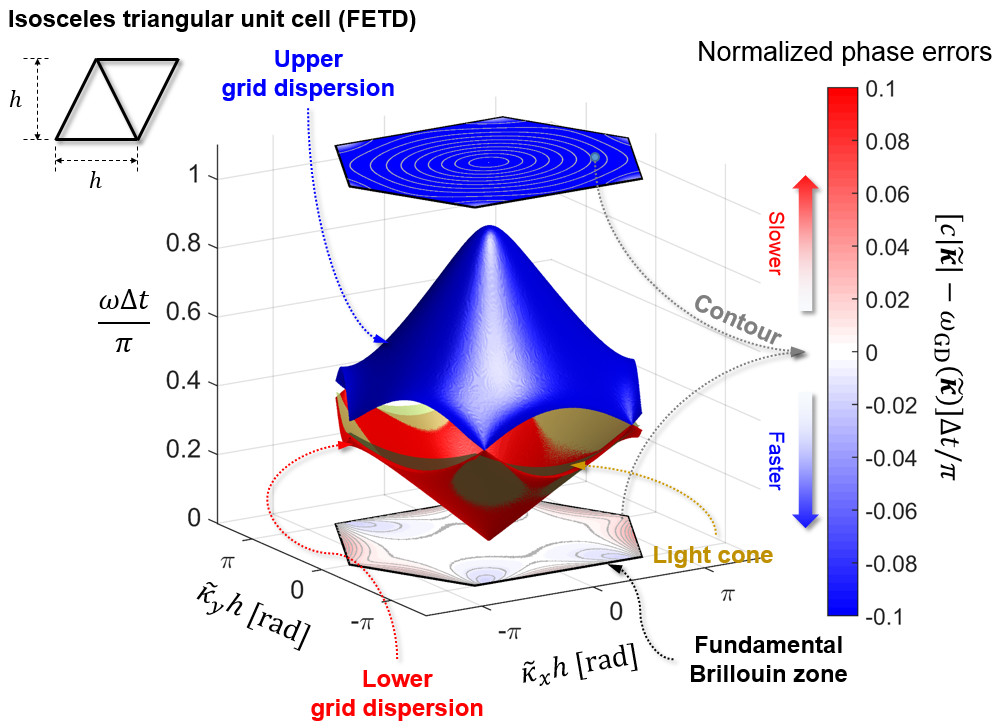}}
\subfloat[\label{fig:gd_fetd_isom_2d}]{\includegraphics[width=2.1in]{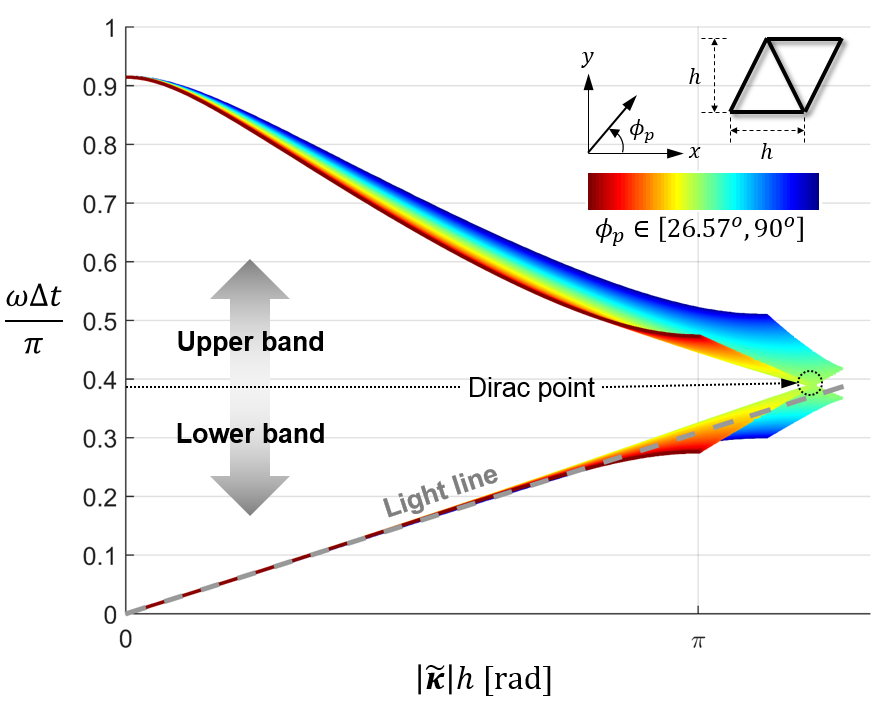}}
\caption{Numerical grid dispersion for the FETD scheme on the ISOT mesh with CFL number equal to one. Unlike the FDTD or FETD-SQ cases, this 
diagram exhibits an additional (upper) dispersion band. (a) 
The red (lower band) and blue (upper band) color surfaces represent the dispersion diagram of the normalized 
frequency $\omega\Delta t/\pi$ versus the normalized numerical wavenumber $\tilde{\boldsymbol{\kappa}}h$ in radians. The olive color surface represents the light cone. The contour levels at the bottom and top represent the normalized phase errors (with respect to the color bar).  (b) Projected dispersion curves for different wave propagation angles with respect to the $x$ axis $\phi_{p}\in\left[26.57^{o},90^{o}\right]$.}
\label{fig:gd_fetd_isom_cfl_1}
\end{figure}
Similar to the RAT case, the numerical dispersion diagram exhibits both lower and upper dispersion bands as shown in Fig. \ref{fig:gd_fetd_isom_cfl_1}.
Fig. \ref{fig:gd_fetd_isom_cfl_1_beam} shows the NCR prediction for a beam bulk velocity $v_{\text{b}}=0.9c$.
There is no NCR caused by the fundamental beam resonances in this case; however, NCR is still excited by waves coupling to aliased beam resonances as illustrated in Fig. \ref{fig:gd_fetd_isom_cfl_1_b_0_9_beam_2d}, where $u \in \left\{-5,-4,...,4,5\right\}$ and $v \in \left\{-3,-2,...,2,3\right\}$.
\begin{figure}
\centering
\subfloat[\label{fig:gd_fetd_isom_cfl_1_b_0_9_beam_3d}]{\includegraphics[width=2.5in]{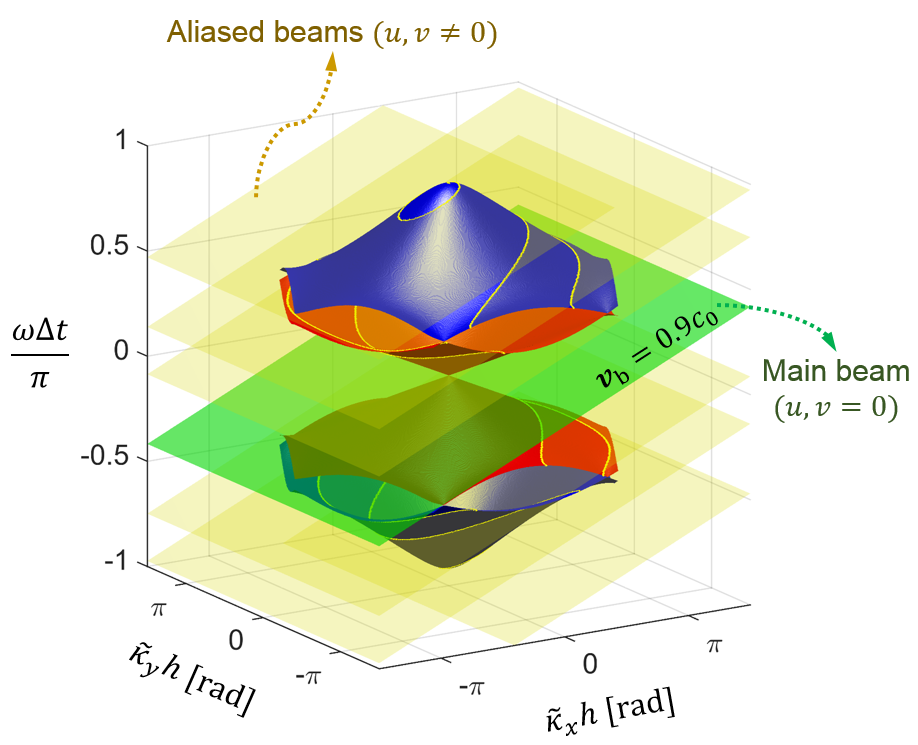}}
~~
\subfloat[\label{fig:gd_fetd_isom_cfl_1_b_0_9_beam_2d}]{\includegraphics[width=2.5in]{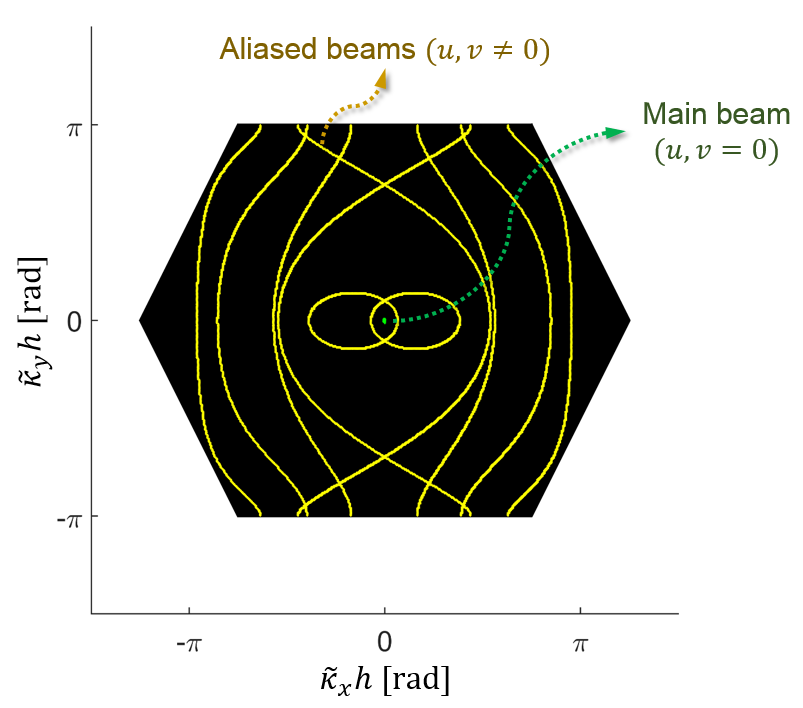}}	
\caption{
Analytic prediction of NCR for the FETD-based EM-PIC scheme on the ISOT mesh assuming a plasma beam with bulk velocity $v_{\text{b}}=0.9c$. (a) Dispersion diagram. (b) NCR solution contours projected onto the first Brillouin zone in the 
 $\tilde{\boldsymbol{\kappa}}$-space.}
\label{fig:gd_fetd_isom_cfl_1_beam}
\end{figure}

\subsubsection{Highly-irregular triangular-element (HIGT) mesh}
In contrast to the SQ, RAT, and ISOT meshes considered above, the HIGT mesh does not have a periodic layout of elements. The aperiodic layout of the elements in the HIGT mesh precludes a similar type of analytical study of NCR as made before for the other meshes. As a result, NCR effects in HIGT are investigated exclusively by means of numerical simulations in the next Section. 
Note that the phase errors due to numerical dispersion on the meshes with periodic triangular elements, as seen in Figs. \ref{fig:gd_fetd_rat_cfl_1} and \ref{fig:gd_fetd_isom_cfl_1}, can be positive or negative (i.e. with dispersion curves above or below the light line) depending on angle of propagation relative to the (local) element orientation.
Therefore, it is expected that for an unstructured (aperiodic) mesh composed of
triangular elements of different shapes and orientations, the {\it cummulative} phase error may be reduced due to some cancellation effects. This result has been numerically observed before in~\cite{wu1997advantages}.

\section{Numerical Experiments}\label{sec:Numerical_Experiment}
In this Section, EM-PIC simulations are conducted to verify the analytic predictions made in the previous Section.
In addition, FETD-based EM-PIC  simulations on the HIGT mesh (for which no analytical prediction is available) are also included for comparison.  
Two basic scenarios are considered here: a relativistically-drifting (electron-positron) pair-plasma and a single electron-positron pair moving in the relativistic regime~\footnote{{\color{black}{
Recently, PIC simulations in the Lorentz boosted frame~\cite{Vayarxiv, Martins_NP}
have been shown to overcome significant scale differences in certain plasma problems~\cite{VayPRL}. PIC simulations in the Lorentz boosted frame can be modeled by replacing background quasi-neutral plasmas in the lab frame by relativistic pair plasmas; however, relativistic pair plasmas can produce significant numerical Cherenkov radiation (NCR).}}}. First, consider a relativistic pair-plasma drifting along the $x$-axis with velocity $v_{\text{b}}=0.9c$ (equivalent to a Lorentz factor of $\gamma_{b}\approx2.3$). The electron plasma frequency is set to $\omega_{pe}\approx4\times10^{5}$ rad/s and the electron density to $n_{e}=1\times10^{8}$ $\text{m}^{-2}$ (same for positrons). Superparticles representing $2.5\times10^{6}$ charged particles are used for each species (electrons and positrons).
The average number of superparticle per cell is set to $40$ for the SQ mesh and $20$ for the RAT, ISOT, HIGT meshes~\footnote{Note that for similar edge sizes, SQ mesh elements are about twice the size of the triangular mesh elements}.
In all cases the problem domain $\Omega=\left\{\left(x,y\right)\in\left[0,128\right]^{2}\right\}$ $\text{m}^{2}$ is terminated by periodic boundary conditions {\color{black}(PBC)} for both fields and particles.
{\color{black}
For field PBC implementation, we set every pair of two edges on the vacuum boundary, which are either horizontally or vertically separated by $128$ m, physically identical. In other words, each pair of two edges shares one DoF and basis function. This results in the reduction of the number of DoFs in the finite-element linear system. It can be easily implemented for periodic layouts due to their regularity. For the HIGT mesh, we created nodes and edges on the vacuum boundary first in such a way that they satisfy the above PBC criteria and then generated interior nodes and edges based on Delaunay triangulation method, consequently, we have the regularity on the vacuum boundary whereas grids are irregular inside the domain.
}
Initially, all superparticles are uniformly distributed over the entire simulation domain and each pair of superparticles is placed at the same position to produce zero net initial fields.
The initial velocity distribution in the beam rest frame  $\mathbf{v}^{\text{bf}}$ for both electrons and positrons is Maxwellian with thermal velocity $v_{\text{th}}=0.005c$ (see Fig. \ref{fig:PHS_bf} and \ref{fig:V_amp_bf}).
\begin{figure}
\centering
\subfloat[\label{fig:PHS_bf}]{\includegraphics[width=1.25in]{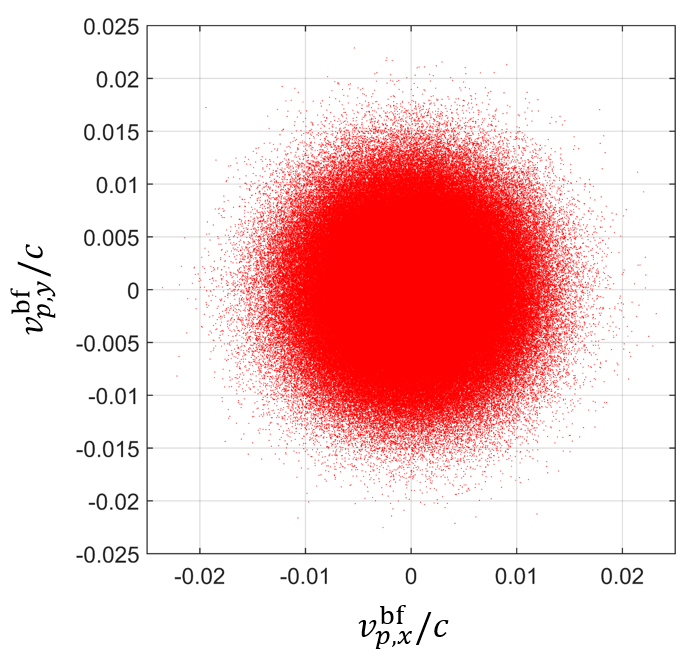}}
\subfloat[\label{fig:V_amp_bf}]{\includegraphics[width=1.2in]{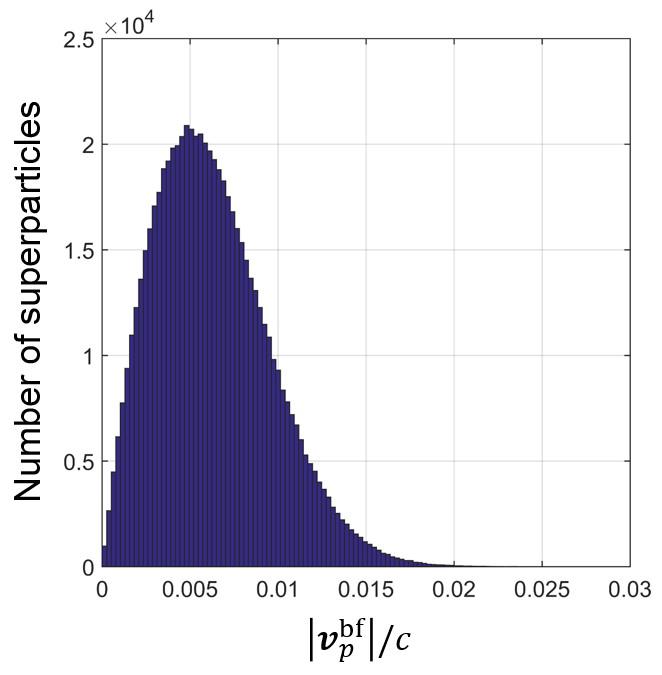}}
\subfloat[\label{fig:PHS_lf}]{\includegraphics[width=0.935in]{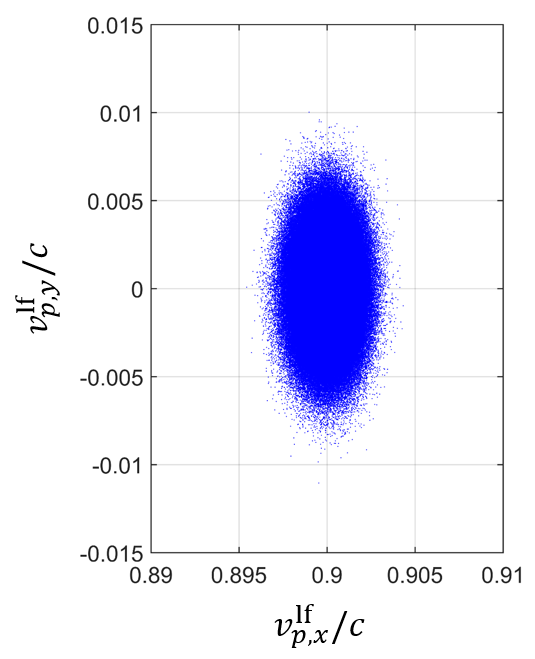}}
\subfloat[\label{fig:V_amp_lf}]{\includegraphics[width=1.2in]{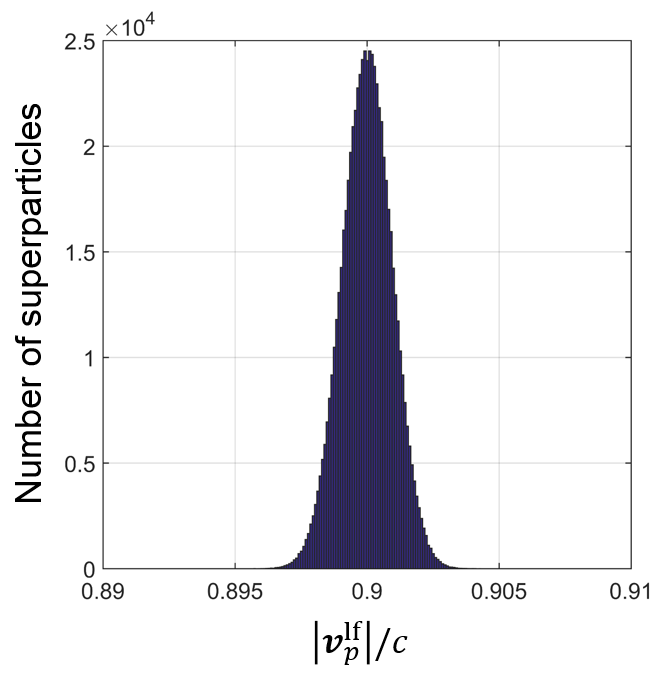}}
\caption{Initial velocity distributions for a relativistic pair plasma beam with bulk velocity $v_{\text{b}}=0.9c$ ($\gamma_{\text{b}}\approx2.3$). (a) Phase space in the beam rest frame. (b) Velocity distribution in the beam rest frame. (c) Phase space in the laboratory frame. (d) Velocity distribution in the laboratory frame.}
\label{fig:beam_info}
\end{figure}
The resulting Debye length equals to $\lambda_{D}^{\text{bf}}=\sqrt{{\epsilon_{0}v_{th}^{2}m_{e}}/\left({n_{e}q_{e}^{2}}\right)}\approx2.657$ m.
The initial velocity distribution in the laboratory frame $\mathbf{v}^{\text{lf}}$ is shown in Fig.~\ref{fig:PHS_lf} and Fig. \ref{fig:V_amp_lf}. 
The
relationship between 
$v_{x}^{\text{lf}}$ and $v_{x}^{\text{bf}}$ can be obtained by applying the Lorentz velocity transformation $v_{x}^{\text{lf}}=\left(v_{x}^{\text{bf}}+v_{b}\right)/\left(1+v_{\text{b}}v_{x}^{\text{bf}}/c^{2}\right)$ and $v_{y}^{\text{lf}}=v_{y}^{\text{bf}}\sqrt{1-\left(v_{\text{b}}/c\right)^{2}}/\left(1+v_{\text{b}}v_{x}^{\text{bf}}/c^{2}\right)$.
Due to length contraction, the Debye length in the laboratory frame reduces to $\lambda_{D}^{\text{lf}}=\lambda_{D}^{\text{bf}}/\gamma_{b}\approx1.158$ m.

EM-PIC simulations are performed based on the following setups: (1-a) FDTD-based solver on the SQ mesh, (1-b) FETD-based solver on the SQ mesh, 
(2) FETD-based solver on the RAT mesh, 
(3) FETD-based solver on the ISOT mesh, 
and (4)  FETD-based solver on the HIGT mesh.
The HIGT mesh is shown in Fig. \ref{fig:pic_hustr_profile}.
Note again that the SQ, RAT, and ISOT meshes have periodic layouts of elements, whereas HIGT has an aperiodic layout.  
To ensure a good mesh quality in the latter case, the angles of triangular elements are enforced to be no less than $30^{o}$. The average angle is near $60^{o}$.
The angle distribution (histogram) is shown in  Fig. \ref{fig:pic_hustr_profile}. 
\begin{figure}
	\centering
	\subfloat[\label{fig:usm_mesh}]{\includegraphics[width=1.47in]{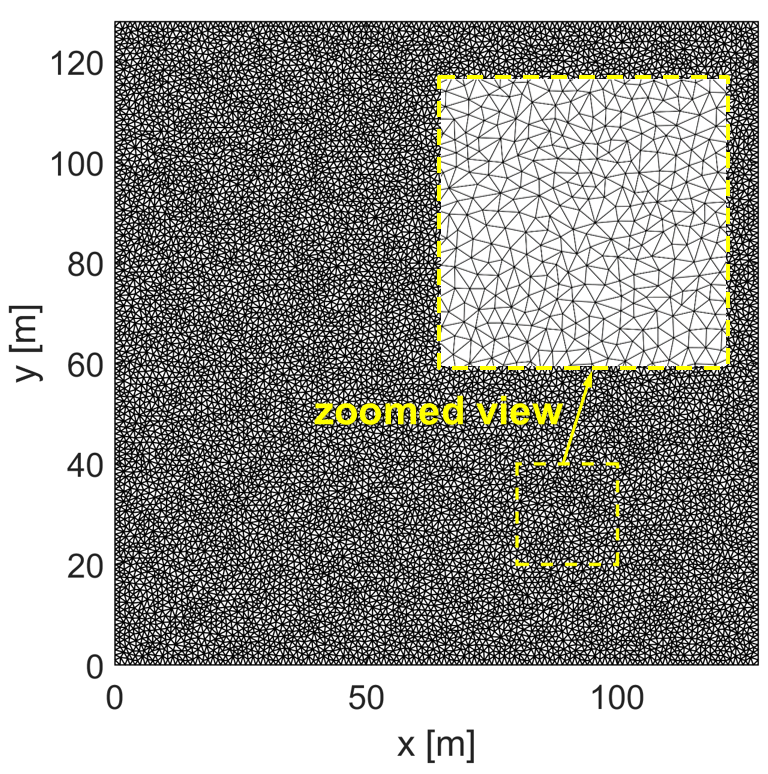}}
	\subfloat[\label{fig:usm_mesh_edge_length}]{\includegraphics[width=1.5in]{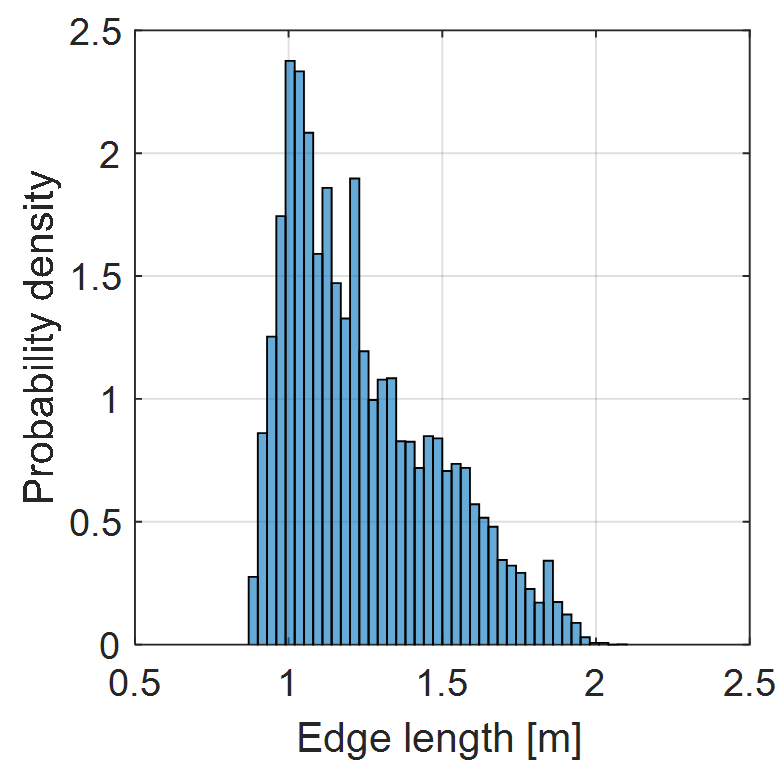}}
	\subfloat[\label{fig:usm_mesh_angle}]{\includegraphics[width=1.5in]{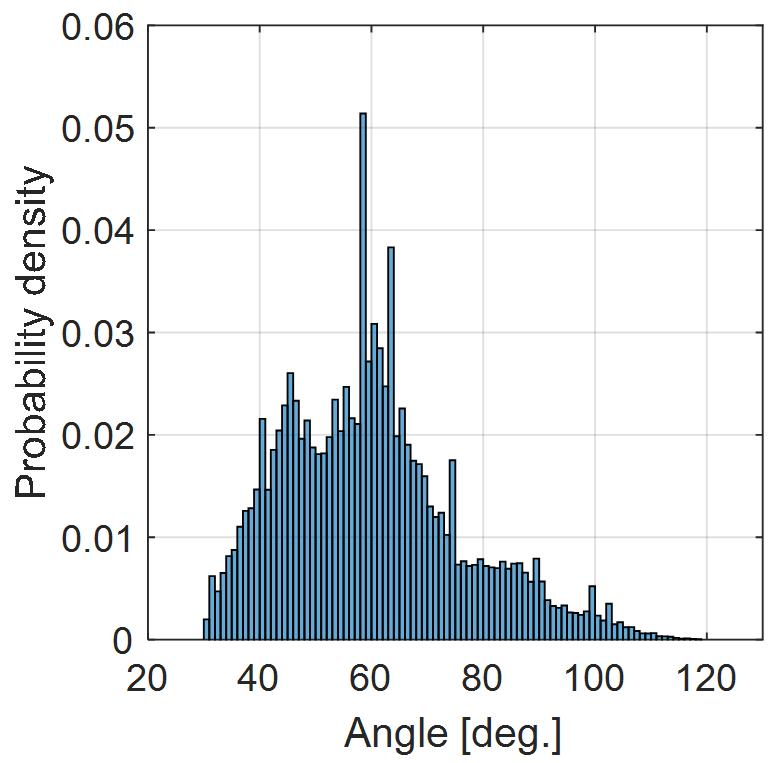}}
	\caption{(a) HIGT mesh. (b) Histogram of the edge lengths. (c) Histogram of the triangular element angles.}
	\label{fig:pic_hustr_profile}
\end{figure}
All meshes are designed so that the average edge length $l_{\text{avg}}$ is comparable to $\lambda_{D}^{\text{lf}}$ to mitigate self-heating effects. 
Table \ref{tab:mesh} lists the basic properties of the four types of mesh.
\begin{table}[h]
\centering
\caption{Basic meshes properties.}
\small
\begin{tabular}{l*{6}{c}r}
{\bf parameters} & {\bf simulation type: (1-a)} & {\bf (1-b)} & {\bf (2)} & {\bf (3)} & {\bf (4)} \\
\hline 
$h$ [m] & \qquad \qquad \qquad 1.00 & 1.00 & 1.00 & 1.00 & -  \\
$l_{\text{avg}}$ [m] & \qquad \qquad \qquad 1.00 & 1.00 & 1.14 & 1.08 & 1.24  \\
$N_{0}$ ($\#$ nodes) & \qquad \qquad \qquad 16,641 & 16,641 & 16,641 & 16,641 & 13,239  \\
$N_{1}$ ($\#$ edges) &\qquad \qquad \qquad  33,024 & 33,024 & 49,408 & 49,408 & 39,202  \\
$N_{2}$ ($\#$ faces) & \qquad \qquad \qquad  16,384 & 16,384 & 32,768 & 32,768 & 25,964  \\
$\Delta t_{\text{max}}$ [ns] & \qquad \qquad \qquad 2.35 & 1.35 & 1.11 & 1.03 & 1.02  \\
\hline
\end{tabular}
\label{tab:mesh}
\end{table}

In the deep relativistic regime (i.e. very large $\gamma_{\text{b}}$) for 2-D FDTD-based EM-PIC simulations, the optimal time step $\Delta t_{\text{mag}}$ for the lowest rate of NCR production has been determined~\cite{godfrey2013numerical,xu2013numerical} to be
$\Delta t_{\text{opt}}\approx0.9192\Delta t_{\text{max,2D}}=0.9192h/\sqrt{2}c$
where $\Delta t_{\text{max,2D}}$ is the maximum time step for stability, as dictated by the CFL condition. 
On the other hand, the NCR growth rate in the mildly relativistic regime has been observed to monotonically decrease as $\Delta t$ increases, that is $\Delta t_{\text{opt}} \rightarrow \Delta t_{\text{max,2D}}$. In either case, $\Delta t_{\text{opt}}$ does not differ substantially from
$\Delta t_{\text{max,2D}}$. The present FDTD-based EM-PIC simulations adopt $\Delta t_{\text{max,2D}}$ as a reference for comparison.

To obtain NCR dispersion maps in the $\tilde{\boldsymbol{\kappa}}$-space representation, the $z$ component of the $\mathbf{B}$ field is measured across the mesh at the end of the simulation ($t = 47~\mu\text{s}$). 
A 2-D fast Fourier transform (FFT) is then performed on the sampled data to obtain $\mathbf{B}$ in the $\tilde{\boldsymbol{\kappa}}$-space representation.
Fig.~\ref{fig:pic_str_b_0_9} shows contour plots of the 
amplitude of $\mathbf{B}$ in log scale over the first Brillouin zone 
in  the $\tilde{\boldsymbol{\kappa}}$-space 
from FDTD-based and FETD-based EM-PIC simulations on the {\color{black}SQ mesh} (cases (1-a) and (1-b)). Fig.~\ref{fig:pic_ustr_b_0_9} illustrates the same for FETD-based EM-PIC simulations on the RAT and ISOT meshes (cases (2) and (3)). 
Analytic prediction curves are superimposed on the simulation results in Figs. \ref{fig:pic_fdtd_b_0_9_comp}, \ref{fig:pic_fetd_qdl_b_0_9_comp}, \ref{fig:pic_fetd_rat_b_0_9_comp} and \ref{fig:pic_fetd_istm_b_0_9_comp}. The black and gray colors denote fundamental and aliased beams, respectively.
A very good agreement is observed between the numerical results and the analytic predictions in all cases. 
\begin{figure}
\centering
\subfloat[\label{fig:pic_fdtd_b_0_9}]{\includegraphics[width=2in]{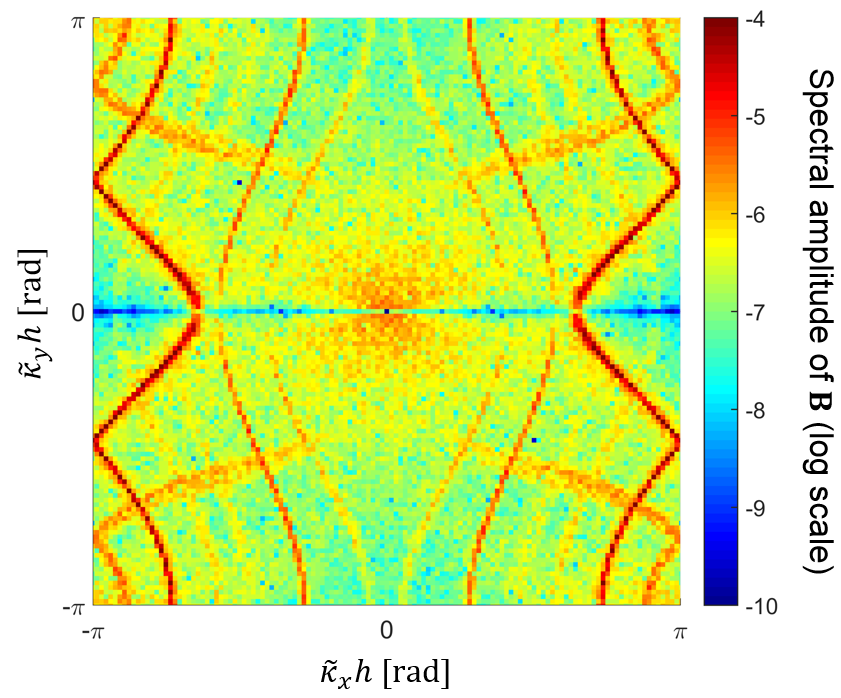}}
~~
\subfloat[\label{fig:pic_fdtd_b_0_9_comp}]{\includegraphics[width=2in]{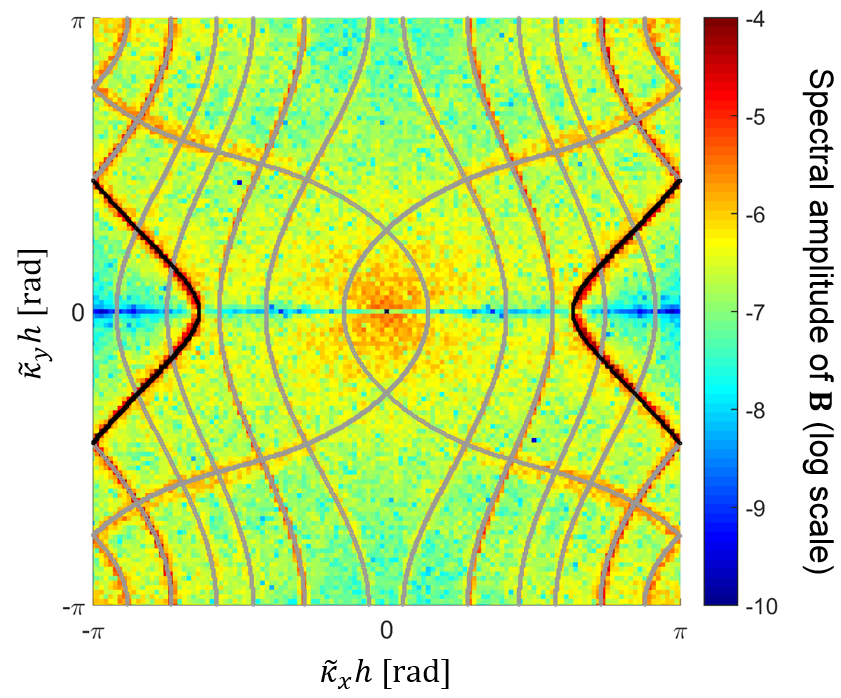}}
\\
\subfloat[\label{fig:pic_fetd_qdl_b_0_9}]{\includegraphics[width=2in]{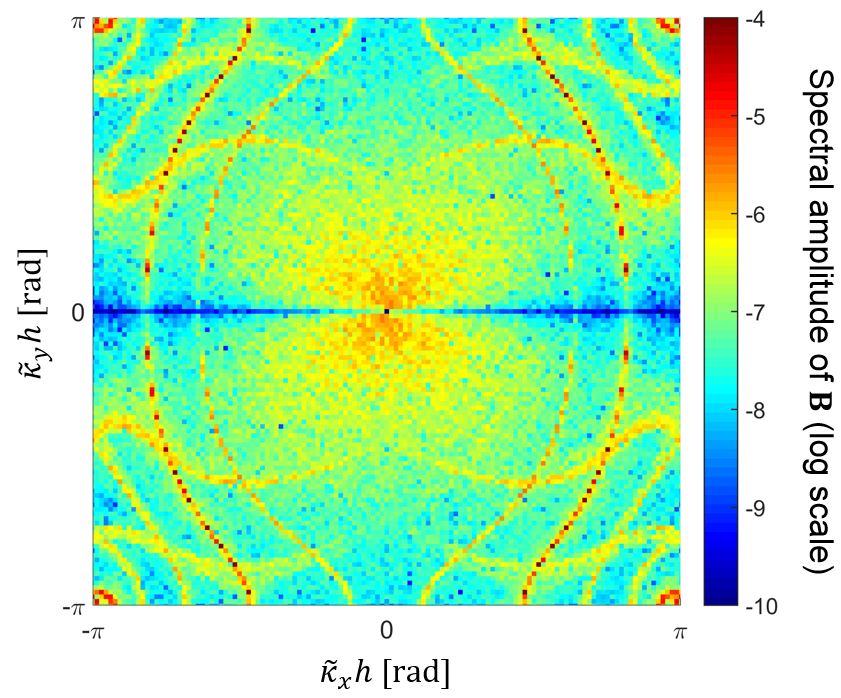}}
~~
\subfloat[\label{fig:pic_fetd_qdl_b_0_9_comp}]{\includegraphics[width=2in]{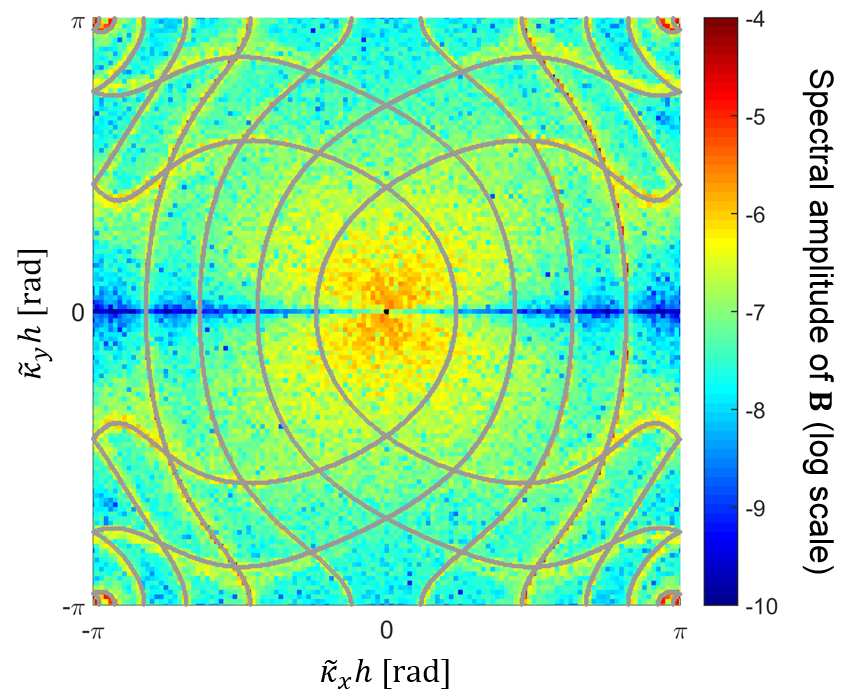}}
\caption{
$\mathbf{B}$ field amplitude distribution (log scale) over the first Brillouin zone in the
 $\tilde{\boldsymbol{\kappa}}$-space as measured from EM-PIC simulation snapshots at $47~\mu\text{s}$. (a) and (c) plots correspond to FDTD- and FETD-based EM-PIC simulations on the SQ mesh, respectively. In (b) and (d), the analytical predictions are superimposed to the numerical results.}
\label{fig:pic_str_b_0_9}
\end{figure}
\begin{figure}
\centering
\subfloat[\label{fig:pic_fetd_rat_b_0_9}]{\includegraphics[width=2in]{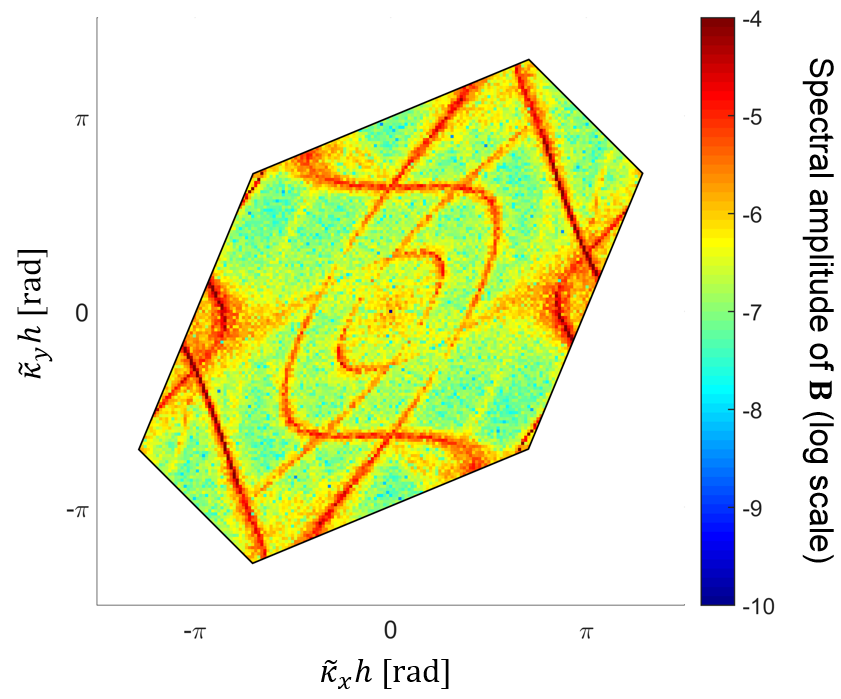}}
~~
\subfloat[\label{fig:pic_fetd_rat_b_0_9_comp}]{\includegraphics[width=2in]{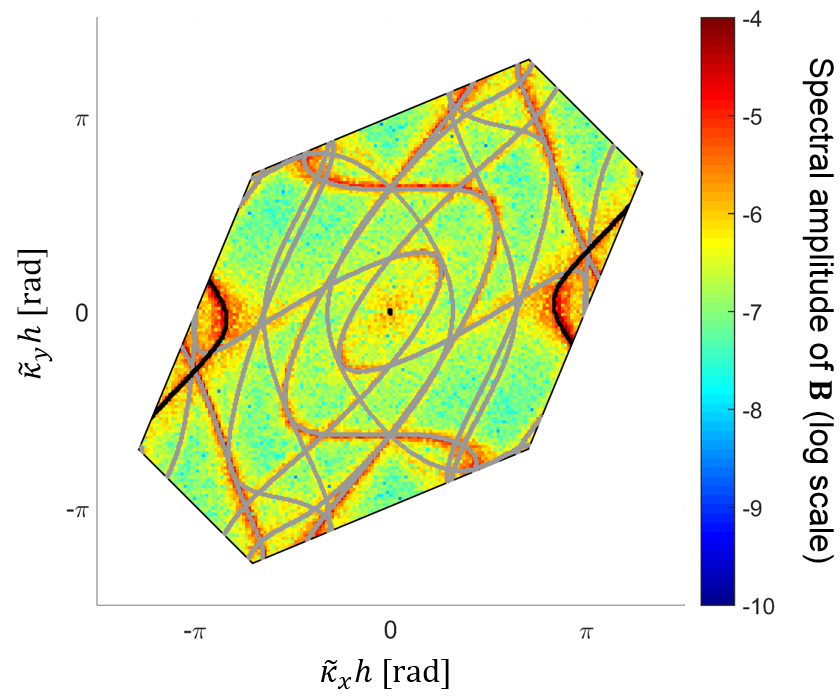}}
\\
\subfloat[\label{fig:pic_fetd_istm_b_0_9}]{\includegraphics[width=2in]{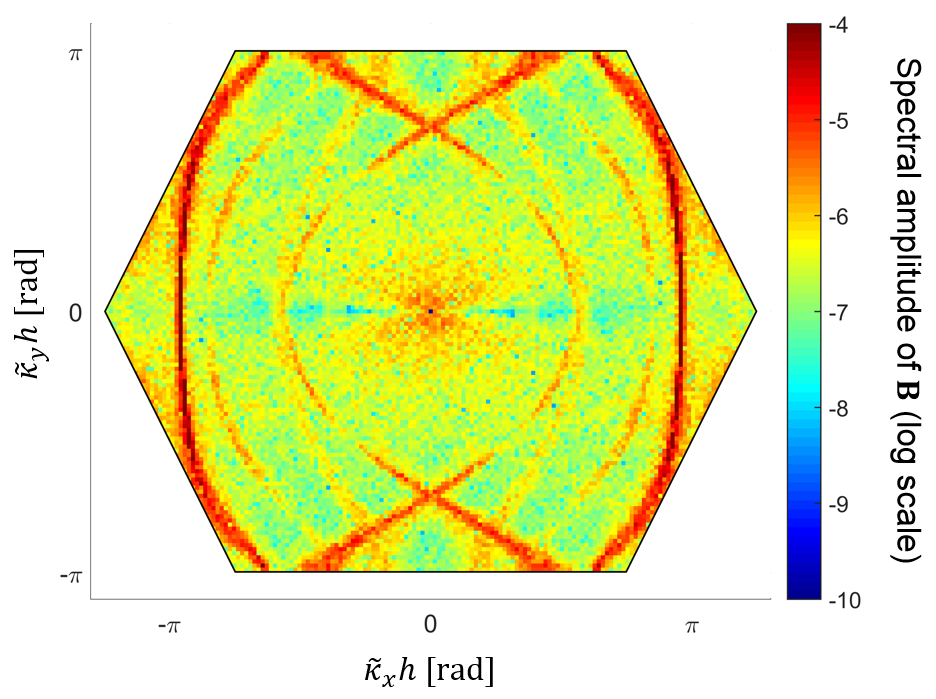}}
~~
\subfloat[\label{fig:pic_fetd_istm_b_0_9_comp}]{\includegraphics[width=2in]{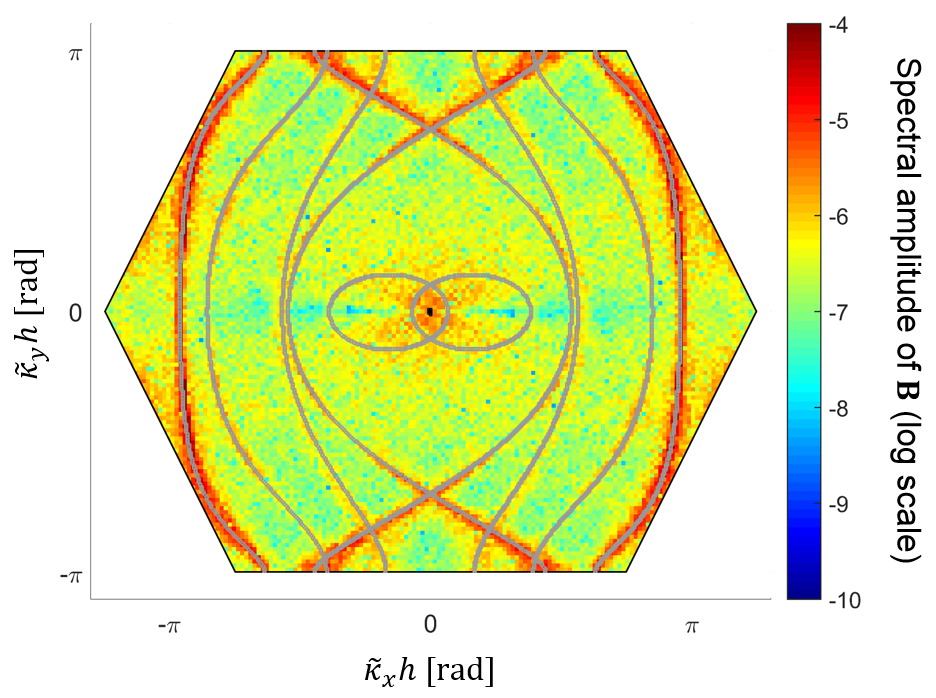}}
\caption{$\mathbf{B}$ field amplitude distribution (log scale) over the first Brillouin zone in the 
 $\tilde{\boldsymbol{\kappa}}$-space as measured from EM-PIC simulation snapshots at $47~\mu\text{s}$. (a) and (c) plots correspond to FETD-based EM-PIC simulations on the RAT and ISOT meshes, respectively. In (b) and (d),  the analytical predictions are superimposed to the numerical results.}
\label{fig:pic_ustr_b_0_9}
\end{figure}

It is observed that on periodic triangular grids, NCR can be purely transverse and longitudinal unlike the SQ case (FDTD or FETD) as discussed in \cite{xu2013numerical}. On the other hand, propagation along certain directions (`characteristics edges') is prohibited.
It is evident also that the NCR distribution in the
 $\tilde{\boldsymbol{\kappa}}$-space is strongly dependent on the mesh element shapes.
The existence of the upper grid dispersion bands is also confirmed from their contributions to the NCR solutions in
the $\tilde{\boldsymbol{\kappa}}$-space.
Compared to the FDTD-based simulation and to the FETD-based simulations on the RAT and ISOT meshes, the NCR observed in FETD-based EM-PIC simulation on the SQ mesh exhibits weaker amplitudes. Also of note is that even though there is no NCR produced by the fundamental beam in
FETD-ISOT case (3), the NCR caused by aliased beams is stronger than in FETD-SQ case (1-b). 


The NCR amplitude distribution in the
 $\tilde{\boldsymbol{\kappa}}$-space 
for the FETD-based EM-PIC simulation on the HIGT mesh is shown in Fig. \ref{fig:pic_hustr_b_0_9}.
Unlike the previous cases, the aperiodic layout of mesh elements of the HIGT mesh precludes spatial coherency. Instead, a diffusive-like (spatially incoherent) pattern in the $\tilde{\boldsymbol{\kappa}}$-space is present.
Fig. \ref{fig:cut_view} shows a quantitative comparison of the $\mathbf{B}$ field amplitude distribution in the
 $\tilde{\boldsymbol{\kappa}}$-space between the FDTD and FETD-HIGT cases.
Fig. \ref{fig:ver_cut} depicts the amplitude of $\mathbf{B}$ versus $\tilde{\kappa}_{y}h$ at some fixed values of $\tilde{\kappa}_{x}h$ and vice-versa in Fig. \ref{fig:hor_cut}.
This corresponds to vertical and horizontal cuts, respectively, on
Figs. \ref{fig:pic_fdtd_b_0_9} and \ref{fig:pic_hustr_b_0_9}.
It can be seen that the peak spectral amplitude in the FDTD case is about two orders of magnitude larger than that in the FETD-HIGT case.
The peaks in the FDTD result correspond to spatially coherent NCR modes. In the FETD-HIGT case, on the other hand, NCR is more evenly spread in 
the $\tilde{\boldsymbol{\kappa}}$-space.
\begin{figure}
\centering
\includegraphics[width=2.in]{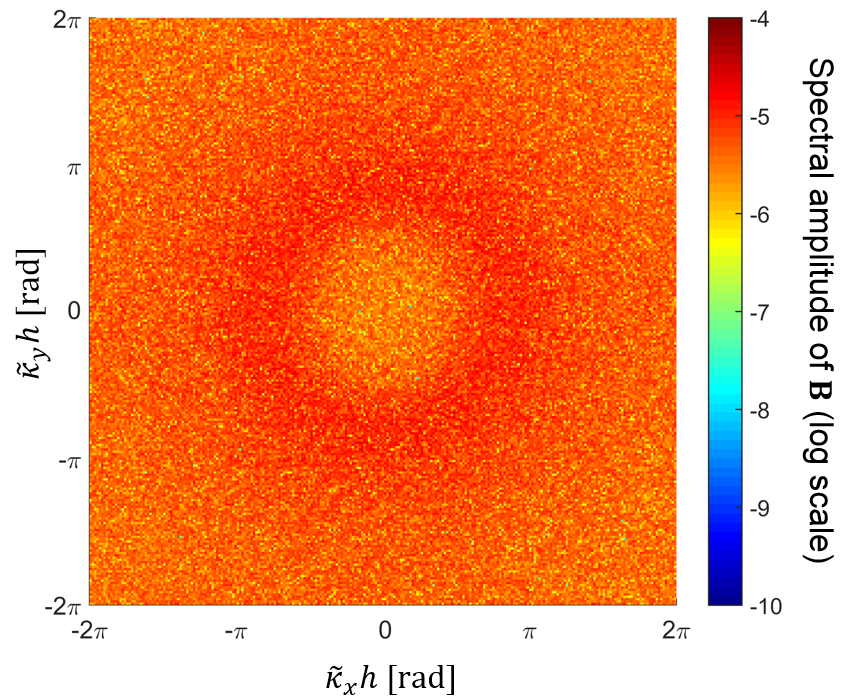}
\caption{$\mathbf{B}$ field amplitude distribution (log scale) over the first Brillouin zone in the
 $\tilde{\boldsymbol{\kappa}}$-space as measured from FETD-based EM-PIC simulation snapshots at $47~\mu\text{s}$ on the HIGT mesh.}
\label{fig:pic_hustr_b_0_9}
\end{figure}
\begin{figure}
\centering
\subfloat[\label{fig:ver_cut}]{\includegraphics[width=2.5in]{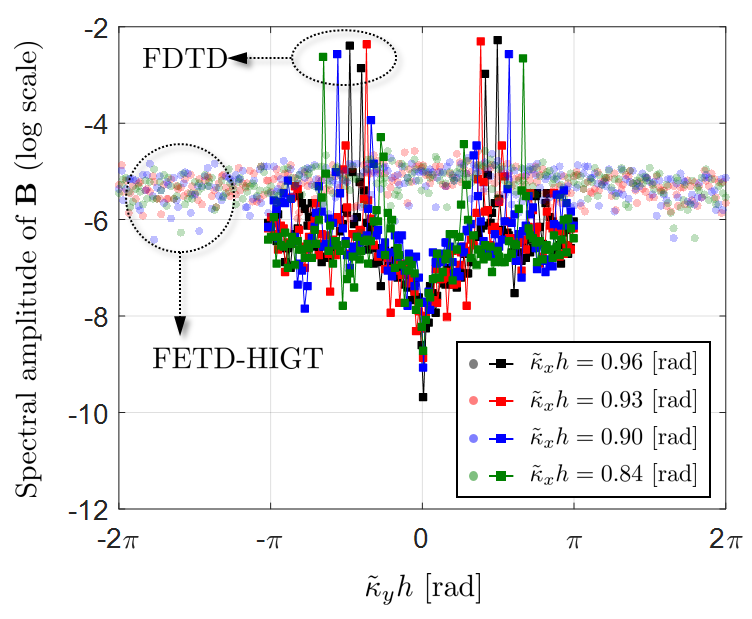}}
\subfloat[\label{fig:hor_cut}]{\includegraphics[width=2.5in]{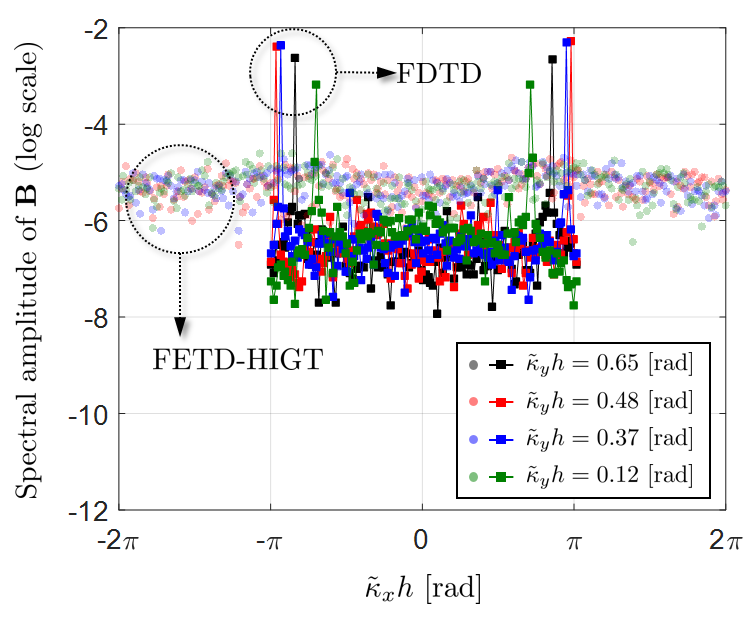}}
\caption{The qualitative comparison of the $\mathbf{B}$ field amplitude distribution (log scale) on the 
 $\tilde{\boldsymbol{\kappa}}$-space between FDTD and FETD-HIGT cases. (a) shows the spectral amplitude of $\mathbf{B}$ versus $\tilde{\kappa}_{y}h$ at some fixed values of $\tilde{\kappa}_{y}h$ and vice-versa in (b).}
\label{fig:cut_view}
\end{figure}

As also observed in~\cite{ikeya2015stability,nuter2016suppressing}, these results above confirm that the observed magnetic field originates from NCR. Similar to the analysis done in~\cite{ikeya2015stability,nuter2016suppressing} we next compare the growth rate of the NCR-induced magnetic field by evaluating the total magnetic field energy on the mesh given by~\cite{teixeira2013differential}:
\begin{flalign}
W_{m}^{n+\frac{1}{2}}=\frac{1}{2}\mathbf{b}^{n+\frac{1}{2}}\cdot\left[\star_{\mu^{-1}}\right]\cdot\mathbf{b}^{n+\frac{1}{2}}.
\end{flalign}
The above expression is computed as a function of time for the various types of mesh considered above. 
These results are shown in Fig. \ref{fig:mag_en}.
Among all cases, the FETD-based EM-PIC simulation on the SQ mesh exhibits the smallest growth rate at earlier times. 
More importantly, the magnetic energy produced by NCR in the SQ, RAT, and ISOT meshes (periodic layouts) reaches saturation levels which are at least one order of magnitude above that in the HIGT mesh (aperiodic layout). 
This could be attributed to the fact that, as noted above, the latter type of mesh does not support spatially coherent NCR modes. These features could be explored to devise possible strategies for NCR mitigation such as, for example, use of hybrid meshes composed of SQ and HIGT elements in different subdomains.

\begin{figure}
\centering
\includegraphics[width=3.in]{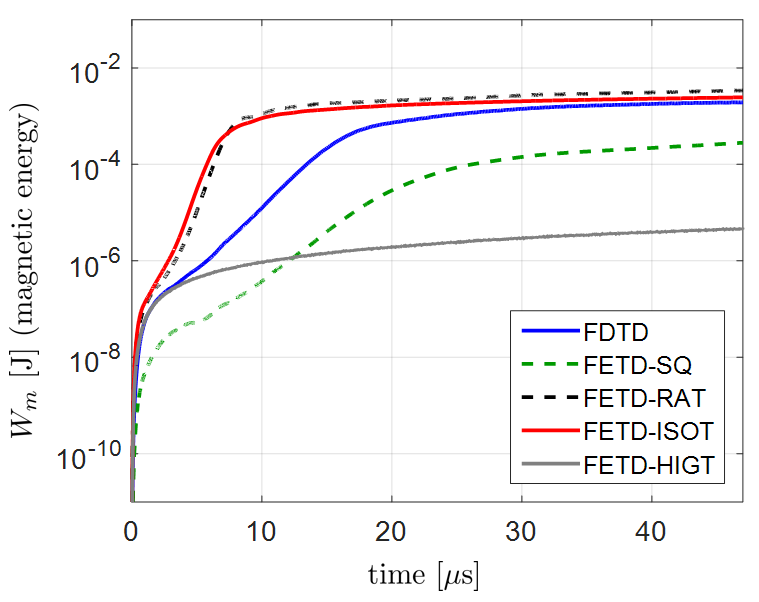}
\caption{Evolution of the magnetic energy $W_{m}$ due to NCR on various meshes.}
\label{fig:mag_en}
\end{figure}

In order to further illustrate the distinct NCR behavior across various meshes, we consider the simulation of a single electron-positron pair moving at relativistic velocity. 
Although strictly speaking an EM-PIC simulation of a single particle pair may not describe very precisely the underlying physics due to the finite mesh resolution, it is nevertheless useful for unveiling NCR patterns. We assume an electron ($e$) and a positron ($p$) are launched with $\mathbf{v}_{e,p}=v_{b}\hat{x} \pm 1.7\times10^{5}\hat{y}$ m/s, respectively, where $v_{b}=0.9c$ m/s. We observe the resulting magnetic field on the very same meshes as considered before. 
Fig. \ref{fig:pic_sp} shows snapshots of magnetic field on each mesh at three time instants, as indicated.
It can be seen that in the case of meshes with periodic layouts, NCR have preferential directions of propagation according to the intersection points in the first Brillouin zone. In contrast, the NCR pattern on the HIGT mesh has a diffusive-like shape originating from the particle trail.
{\color{black} Note that, by Lorentz invariance, any charged particle moving with a constant speed in vacuum should not emit electromagnetic radiation in the lab frame. In addition, in the moving frame there are only static electric fields and no magnetic fields. In the lab frame there are no sources for radiative magnetic fields; instead the magnetic fields are tightly attached to charged particles. This implies that the magnetic energy must be constant in time, hence, the spurious magnetic energy growth in these simulations is a direct evidence of NCR.} 

\begin{figure}
\centering
\subfloat[\label{fig:pic_sp_fdtd_2}]{\includegraphics[width=1.55in]{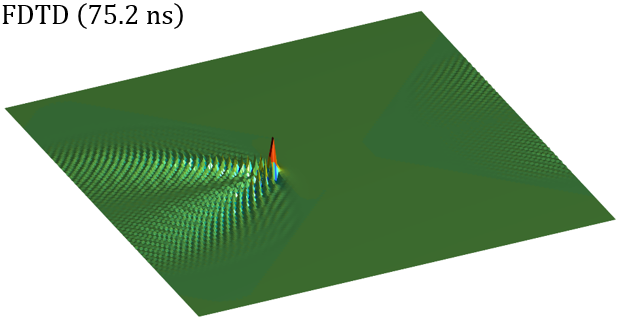}}
\subfloat[\label{fig:pic_sp_fdtd_3}]{\includegraphics[width=1.55in]{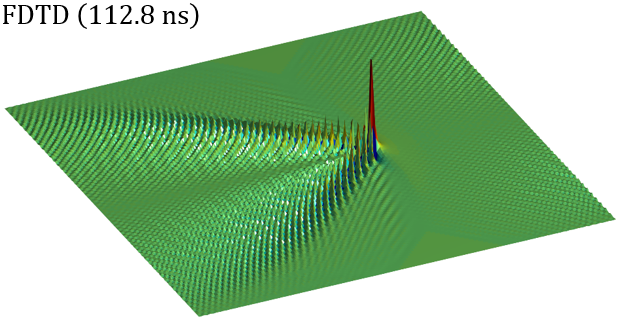}}
\subfloat[\label{fig:pic_sp_fdtd_4}]{\includegraphics[width=1.55in]{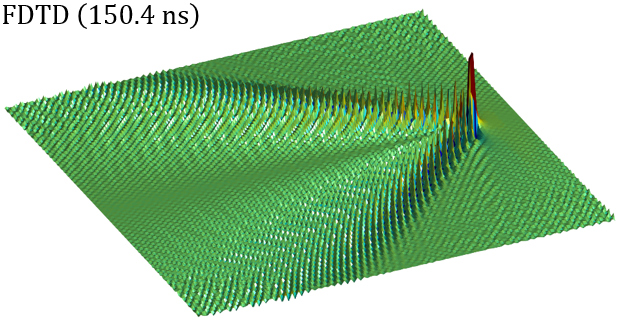}}
\\
\subfloat[\label{fig:pic_sp_fetd_qdl_2}]{\includegraphics[width=1.55in]{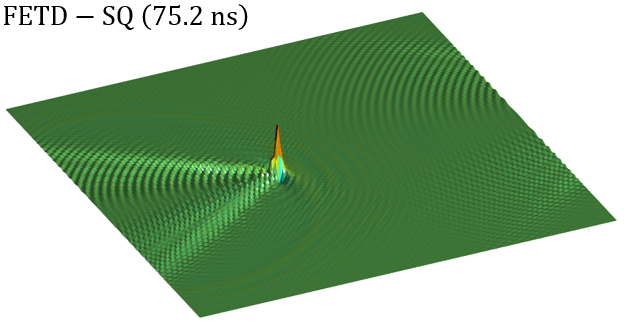}}
\subfloat[\label{fig:pic_sp_fetd_qdl_3}]{\includegraphics[width=1.55in]{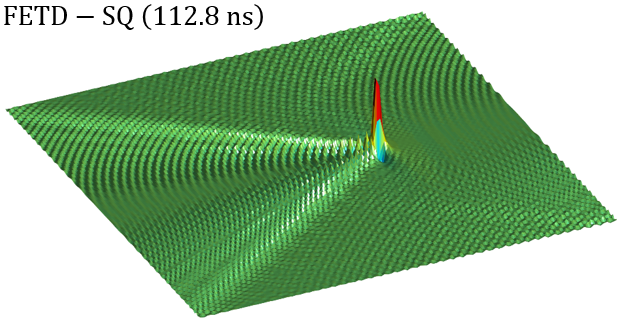}}
\subfloat[\label{fig:pic_sp_fetd_qdl_4}]{\includegraphics[width=1.55in]{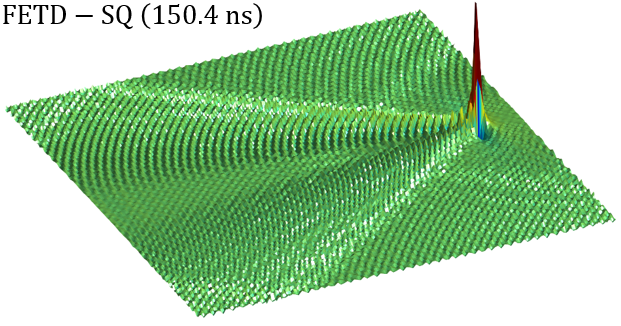}}
\\
\subfloat[\label{fig:pic_sp_fetd_rat_2}]{\includegraphics[width=1.55in]{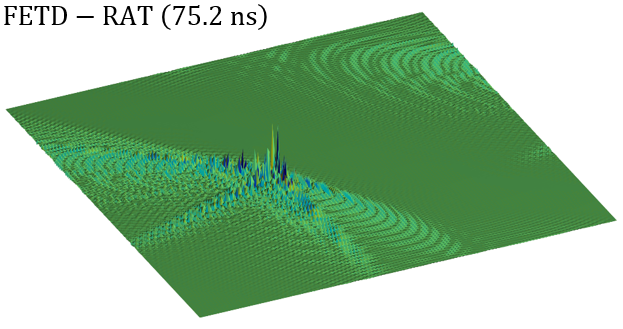}}
\subfloat[\label{fig:pic_sp_fetd_rat_3}]{\includegraphics[width=1.55in]{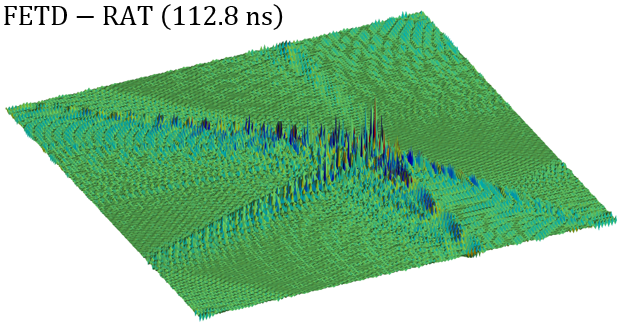}}
\subfloat[\label{fig:pic_sp_fetd_rat_4}]{\includegraphics[width=1.55in]{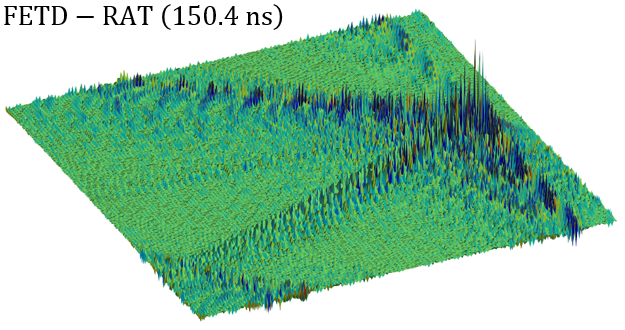}}
\\
\subfloat[\label{fig:pic_sp_fetd_isot_2}]{\includegraphics[width=1.55in]{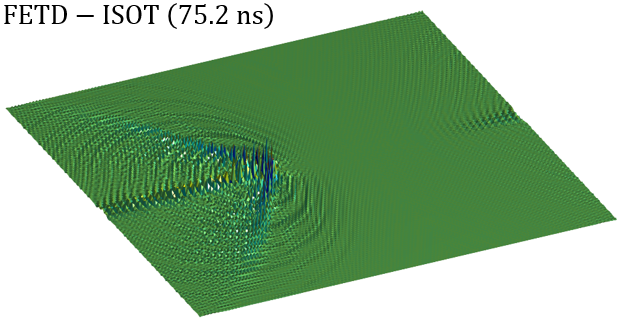}}
\subfloat[\label{fig:pic_sp_fetd_isot_3}]{\includegraphics[width=1.55in]{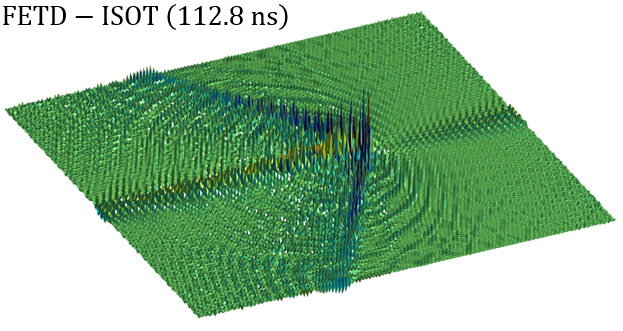}}
\subfloat[\label{fig:pic_sp_fetd_isot_4}]{\includegraphics[width=1.55in]{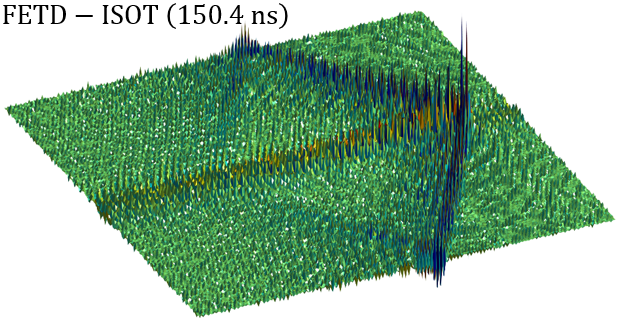}}
\\
\subfloat[\label{fig:pic_sp_fetd_usm_2}]{\includegraphics[width=1.5in]{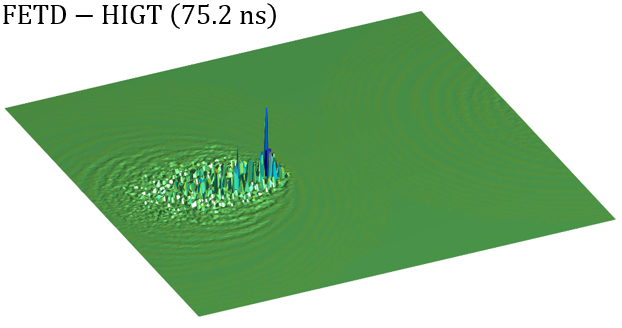}}
\subfloat[\label{fig:pic_sp_fetd_usm_3}]{\includegraphics[width=1.5in]{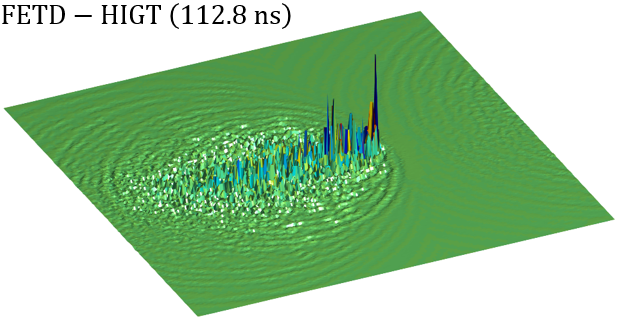}}
\subfloat[\label{fig:pic_sp_fetd_usm_4}]{\includegraphics[width=1.5in]{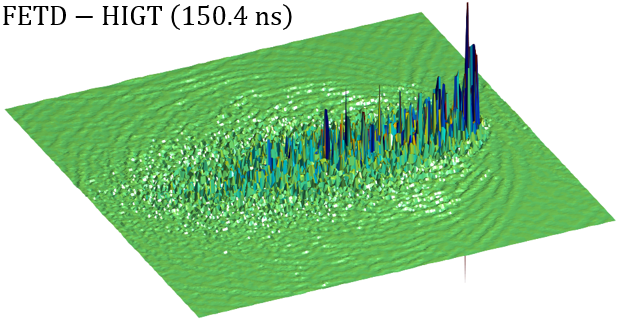}}
\caption{Snaphots  of the magnetic field distribution resulting from EM-PIC simulations of a single electron-positron pair moving relativistically. The snapshots are taken at $75.2$ ns, $112.8$ ns, and $150.4$ ns, as indicated. The results correspond to: (a-c) FDTD-based EM-PIC simulation on SQ mesh , (d-f) FETD-based EM-PIC simulation on SQ mesh, (g-i)  FETD-based EM-PIC simulation on the RAT mesh, (j-l)  FETD-based EM-PIC simulation on ISOT mesh, (m-o)  FETD-based EM-PIC simulation on HIGT mesh.}
\label{fig:pic_sp}
\end{figure}

\section{Concluding Remarks}\label{sec:Conclusion}
We analyzed numerical Cherenkov radiation (NCR) effects produced by finite-element-based EM-PIC simulations on different types of mesh.
Complete dispersion diagrams over the first Brillouin zone were derived for periodic meshes with different element shapes and layouts. Analytical NCR predictions were compared against numerical results from EM-PIC simulations.
Considering a relativistic plasma beam simulation, it was observed that the mesh element shape and mesh layout have a marked influence on the ensuing NCR properties. 
In particular, it was also observed that EM-PIC simulations on an unstructured mesh (with irregular triangular elements) does not support 
spatially coherent NCR modes due to the aperiodic nature of the mesh layout.  In this case, a diffusive-like behavior is observed for the NCR in the spatial domain. Importantly, it was observed that the spurious energy produced by NCR on the unstructured mesh reaches saturation levels that are considerably lower than those on meshes based on periodic layout of (rectangular or triangular) elements. 
For simplicity, the analysis was carried out here in 2-D but it is expected that similar conclusions should apply to 3-D as well. 

{\color{black}
Finally, it should be stressed that the objective of the present work was to provide a diagnosis and comparison of NCR and aliasing effects across different mesh shapes and not across different shape functions. 
Results based on first order Whitney basis functions and without the use of high-order shape functions unveil the baseline NCR effect for a fair comparative analysis to be made across different meshes.
It is expected that the use of high-order shape functions should mitigate NCR effects. For that, the development of high-order shape functions on unstructured grids is an important line of future enquiry.
}


\section*{Acknowledgment}
This work was supported by the Department of Defense, Defense Threat Reduction Agency grant HDTRA1-18-1-0050,  National Science Foundation grant ECCS-1305838,  S\~{a}o Paulo State Research Support Foundation (FAPESP) grant 2015/50268-5, Ohio Supercomputer Center grants PAS-0061 and PAS-0110, and the Ohio State University Presidential Fellowship program.

The content of the information does not necessarily reflect the position or the policy of the U.S. federal government, and no official endorsement should be inferred.

\clearpage
\appendix
\section{FETD Maxwell Field Solver}\label{sec:FETD_Derivation}

The fundamentals of the present finite element time-domain (FETD) Maxwell field solver are briefly summarized in this appendix. 
A more comprehensive discussion of the Maxwell field solver, together with various details on the scatter, gather, and pusher steps of the EM-PIC algorithm can be found in~\cite{moon2015exact,na2016local,na2018relativistic}.

The present FETD field solver is based on the exterior calculus  representation of Maxwell's equations~\cite{burke1985, flanders1989,teixeira1999lattice,kotiuga2004}, and on expressing the field unknowns in terms of discrete differential forms, also known as Whitney forms~\cite{he2007differential,teixeira2014lattice}. 
On a general mesh, the electric field intensity (1-form) and the magnetic flux density (2-form) can be expanded using a linear combination of Whitney 1-forms $w^{(1)}_{j}$ associated with edges (1-cells) $j=1,\ldots,N_1$ and Whitney 2-forms $w^{(2)}_{k}$ associated with facets (2-cells) $k=1,\ldots,N_2$, 
as:
\begin{flalign}
\mathcal{E}\left(t,\mathbf{r}\right)=\sum_{j=1}^{N_{1}} e_{j}\left(t\right) w^{(1)}_{j}\left(\mathbf{r}\right),
\label{eq:e_wh_1_df}
\\
\mathcal{B}\left(t,\mathbf{r}\right)=\sum_{k=1}^{N_{2}} b_{k}\left(t\right) w^{(2)}_{k}\left(\mathbf{r}\right),
\label{eq:b_wh_2_df}
\end{flalign}
where $N_{1}$ and $N_{2}$ are the total number of edges and facets in the mesh, respectively, $e_{j}\left(t\right)$ and $b_{k}\left(t\right)$ represent the discrete DoFs for the electric field intensity and magnetic flux density, respectively. 
Their vector proxies of the above expressions are given by
\begin{flalign}
\mathbf{E}\left(t,\mathbf{r}\right)=\sum_{j=1}^{N_{1}} e_{j}\left(t\right) \mathbf{W}^{(1)}_{j}\left(\mathbf{r}\right),
\label{eq:e_wh_1_vec}
\\
\mathbf{B}\left(t,\mathbf{r}\right)=\sum_{k=1}^{N_{2}} b_{k}\left(t\right) \mathbf{W}^{(2)}_{k}\left(\mathbf{r}\right).
\label{eq:b_wh_2_vec}
\end{flalign}
Explicit expressions for $w^{(1)}_{j}$, $w^{(2)}_{k}$, $\mathbf{W}^{(1)}_{j}$, and $\mathbf{W}^{(2)}_{k}$, can be found in~\cite{teixeira1999lattice}.

By substituting (\ref{eq:e_wh_1_df}) and (\ref{eq:b_wh_2_df}) into Faraday's law, applying the Generalized Stokes' Theorem~\cite{teixeira1999lattice,teixeira2014lattice,cairns1936the}, and using a leap-frog time integration scheme, a discrete representation of Faraday's law on a general mesh is obtained as
\begin{flalign}
\mathbf{b}^{n+\frac{1}{2}}=\mathbf{b}^{n-\frac{1}{2}}-\Delta t \mathbf{C} \cdot \mathbf{e}^{n},
\label{eqn:disc_far}
\end{flalign}
where the superscript $n$ denotes the time step index, $\mathbf{e}=\left[e_{1},e_{2},...,e_{N_{1}}\right]^{T}$, $\mathbf{b}=\left[b_{1},b_{2},...,b_{N_{2}}\right]^{T}$, and $\mathbf{C}$ is an incidence matrix encoding the discrete exterior derivative on the mesh~\cite{teixeira1999lattice,teixeira2014lattice,hughes1981lagrangian,guth1980existence}.
The elements of $\mathbf{C}$  are in the set $\left\{-1,0,1\right\}$.

By a similar process, the discrete representation for Ampere's law takes the form of
\begin{flalign}
\left[\star_{\epsilon}\right]\cdot \mathbf{e}^{n+1}=\left[\star_{\epsilon}\right]\cdot\mathbf{e}^{n}+\Delta t {\mathbf{C}}^{T}\cdot\left[\star_{\mu^{-1}}\right] \cdot \mathbf{b}^{n+\frac{1}{2}},
\label{eq:dal_prime}
\end{flalign}
where $\left[\star_{\epsilon}\right]$ and $\left[\star_{\mu^{-1}}\right]$ are discrete Hodge operators (square positive-definite matrices) encoding both the constitutive properties of the background medium and metric information of the mesh~\cite{teixeira1999lattice,teixeira2014lattice,he2007differential,kim2011parallel,tarhasaari1999some,gillette2011dual,he2006geometric}.
The elements of $\left[\star_{\epsilon}\right]$ and $\left[\star_{\mu^{-1}}\right]$ can be computed by the integrals~\cite{teixeira2014lattice,kim2011parallel,tarhasaari1999some,he2006geometric}.
\begin{flalign}
\left[\star_{\epsilon}\right]_{J,j}&=\int_{\Omega} w_{J}^{(1)}\wedge\left(\star_{\epsilon}{w_{j}^{(1)}}\right)
=\int_{\Omega} \epsilon_{0} w_{J}^{(1)}\wedge\star{w_{j}^{(1)}}
\nonumber\\
&=\underbrace{ 
	\int_{\Omega}\epsilon_{0}\mathbf{W}_{J}^{(1)} \cdot \mathbf{W}_{j}^{(1)} dV
}_{\text{vector proxy representation}},
\\
\left[\star_{\mu^{-1}}\right]_{K,k}&=\int_{\Omega} w_{K}^{(2)}\wedge\left(\star_{\mu^{-1}}{w_{k}^{(2)}}\right)
=\int_{\Omega} \mu_{0}^{-1} w_{K}^{(2)}\wedge\star{w_{k}^{(2)}}
\nonumber\\
&=\underbrace{ 
	\int_{\Omega}\mu_{0}^{-1}\mathbf{W}_{K}^{(2)} \cdot \mathbf{W}_{k}^{(2)} dV
}_{\text{vector proxy representation}},
\end{flalign}
and $\Omega$ is the problem domain.

Both $\left[\star_{\epsilon}\right]$ and $\left[\star_{\mu^{-1}}\right]$ are sparse, positive-definite, and diagonally-dominant square matrices.
Thus, the linear solve in (\ref{eq:dal_prime})
can be performed very quickly.
Nevertheless, since the linear solve is needed at every step $n$ of the time evolution, a sparse approximate inverse (SPAI) of $\left[\star_{\epsilon}\right]$ may be computed priori to the start of the time-update to obviate the need for the linear solve~\cite{kim2011parallel,na2016local}.

{\color{black}
Equation (\ref{eq:st_disc_vec_wave_eq}) represents the space and time discretization of the electric field wave equation, which can be also obtained directly from the discrete form of Maxwell's curl equations, (\ref{eqn:disc_far}) into (\ref{eq:dal_prime}). By substituting (\ref{eqn:disc_far}) into (\ref{eq:dal_prime}) we obtain
\begin{flalign}
\left[\star_{\epsilon}\right]\cdot \mathbf{e}^{n+1}
=
\left[\star_{\epsilon}\right]\cdot \mathbf{e}^{n}
+\Delta t {\mathbf{C}}^{T}\cdot\left[\star_{\mu^{-1}}\right] \cdot \mathbf{b}^{n-\frac{1}{2}}
-\Delta t^{2} {\mathbf{C}}^{T}\cdot\left[\star_{\mu^{-1}}\right] \cdot \mathbf{C}\cdot \mathbf{e}^{n}.
\end{flalign}
Since 
\begin{flalign}
\Delta t {\mathbf{C}}^{T}\cdot\left[\star_{\mu^{-1}}\right] \cdot \mathbf{b}^{n-\frac{1}{2}}=
\left[\star_{\epsilon}\right]\cdot \mathbf{e}^{n}-\left[\star_{\epsilon}\right]\cdot\mathbf{e}^{n-1},
\end{flalign}
we can easily obtain (\ref{eq:st_disc_vec_wave_eq}).
}

{\color{black}
\section{Dispersion Relation of Space Charge Waves on Lattice}\label{sec:beam_dispersion}
Dispersion relation of space charge waves uncoupled to EM waves can be derived by using perturbation theory \cite{basu2015engineering}.\
For simplicity, we consider the one-dimensional (scalar) case where a cold electron beam with a charge density $\rho_{0}$ and (bulk) velocity $v_{0}$ travels in free space along the $x$-axis.
In the absence of magnetic fields, the longitudinal small perturbation imposed to the electron beam modifies the charge density and velocity 
\begin{flalign}
\rho\left(x,t\right)=\rho_{0}\left(x,t\right)+\rho_{1}\left(x,t\right)
\\
v\left(x,t\right)=v_{0}\left(x,t\right)+v_{1}\left(x,t\right)
\end{flalign}
where subscript ${1}$ is used for all perturbed quantities whereas subscript $0$ is for the unperturbed.
Note that $v_{0}$ is equivalent to the beam bulk velocity $v_{\text{b}}$.
Since $J=\rho v$, we can deduce the relationship between the charge density and velocity from the continuity equation by ignoring any products of two perturbed quantities
\begin{flalign}
\left[\frac{\partial}{\partial t}+v_{0}\frac{\partial }{\partial x}\right] \rho_{1}=
D\rho_{1}
=
-\rho_{0}\frac{\partial v_{1}}{\partial x}
\label{eqn:rho_v_rel_pert}
\end{flalign}
where ${\cal D}=\left[\frac{\partial}{\partial t}+v_{0}\frac{\partial }{\partial x}\right]$.
Similarly, the Lorentz force by the small perturbation can be written as
\begin{flalign}
\frac{dv_{1}}{dt}=
{\cal D} v_{1}
=
\frac{e}{m}E_{1}.
\label{eqn:lorentz_force_pert}
\end{flalign}
Taking $\cal D$ to (\ref{eqn:rho_v_rel_pert}) and plugging (\ref{eqn:lorentz_force_pert}) into the resultant equation, we obtain
\begin{flalign}
{\cal D}^{2}\rho_{1}=-\omega_{p}^{2}\rho_{1}
\end{flalign}
where $\omega_{p}=\sqrt{\left(|q_{e}\rho_{0}|\right)/\left(|m_{e}|\epsilon_{0}\right)}$, hence, ${\cal D}=\pm i\omega_{p}$.
Note that Gauss law fixes the relation between $E_{1}$ and $\rho_{1}$.
All the perturbed quantities associated with Maxwell dynamic variables in the free space are proportional to $\text{exp}\left(\tilde{\kappa}_{x}x-i\omega t\right)$ if the time convention $e^{-i\omega t}$ is taken.
Therefore, in Fourier space ${\cal D}=-i \omega+i v_{0}\tilde{\kappa}_{x}$, as a result, dispersion relations for fast and slow space charge waves are given by
\begin{flalign}
\omega=v_{0}\tilde{\kappa}_{x} \pm \omega_{p}.
\end{flalign}
In the limit of $\omega_{p}\rightarrow 0$, it recovers a so called beam line
\begin{flalign}
\omega=v_{0}\tilde{\kappa}_{x}.
\end{flalign}

The above dispersion relation, however, should be modified on the grid (lattice) where the spatial and temporal periodicity are present as $h$ and $\Delta t$, respectively.
In this case, all perturbed quantities are proportional to Floquet modes, which are general solutions of wave equations in the presence of the periodicity, viz. 
\begin{flalign}
f_{1} \propto \sum_{u,v}\text{exp}\left[i\left(\tilde{\kappa}_{x}-\frac{2\pi}{h}u\right)x-i\left(\omega-\frac{2\pi}{\Delta t}v\right)t\right].
\end{flalign}
for integers $u$ and $v$ and where $f_{1}$ denotes the perturbed quantity.
Hence, the resulting beam line on the grid takes the form
\begin{flalign}
\omega-v\frac{2\pi}{\Delta t}=v_{0}\left(\tilde{\kappa}_{x}--\frac{2\pi}{h}u\right).
\end{flalign}
}

{\color{black}
\section{Particle and Grid Interactions}
PIC algorithms are designed to update kinetic parameters (i.e. position, velocity, and force) of superparticles in response to Lorentz force in ambient space.
On the other hand, Maxwell dynamic variables (i.e. field and source) are advanced by electric current density over grids.
Such particle- and grid-associated quantities interact via Gather and Scatter steps which properly connect to one another.

In this work, we employ Whitney-form-based gather/scatter algorithms \cite{moon2015exact} which guarantee the exact charge conservation on irregular grids, from the first principle, within the limit of the machine precision.
Note that each superparticle is modeled as being point-like, implying that the shape function of superparticles is assumed to be a delta function.

\subsection{Gather}
In the gather step, ambient field values are interpolated at the location of particles from discrete field solutions via Whitney forms.
Hence, we can mathematically represent the ambient field values for $p$\textsuperscript{th} particle 
\begin{flalign}
\mathbf{E}\left(\mathbf{r}_{p}^{n},n\Delta t\right)
&\equiv \mathbf{E}_{p}^{n} 
=
\sum_{j=1}^{N_{1}}
e_{j}^{n}\mathbf{W}_{j}^{(1)}\left(\mathbf{r}_{p}^{n}\right)
\\
\mathbf{B}\left(\mathbf{r}_{p}^{n},n\Delta t\right)
&\equiv \mathbf{B}_{p}^{n} 
=
\sum_{k=1}^{N_{2}}
\frac{b_{k}^{n-\frac{1}{2}}+b_{k}^{n+\frac{1}{2}}}{2}
\mathbf{W}_{k}^{(2)}\left(\mathbf{r}_{p}^{n}\right)
\end{flalign}
where $\mathbf{r}_{p}^{n}$ is position of $p$\textsuperscript{th} particle at $n$\textsuperscript{th} time step.
These field values are used to evaluate Lorentz force acting on charged particles in particle pusher step. 

\subsection{Scatter}
The motion of superparticles during $\Delta t$ is converted into the form of electric currents over grids for the subsequent use of field solvers.
Grid currents and charges are evaluated based on Whitney 1- and 0-forms, as
\begin{flalign}
j_{i}^{n+\frac{1}{2}}
&=
\sum_{p=1}^{N_{p}}
\frac{e}{\Delta t}
\int_{\mathbf{r}_{p}^{n}}^{\mathbf{r}_{p}^{n+1}}
\mathbf{W}_{i}^{(1)}\left(\mathbf{r}\right)\cdot d\mathbf{r}
\\
q_{i}^{n}
&=
\sum_{p=1}^{N_{p}}
e W_{i}^{(0)}\left(\mathbf{r}_{p}^{n}\right)
\end{flalign}
where $e$ is electron charge [C] and $j_{i}$ and $q_{i}$ are DoFs of electric current and charge for $i$\textsuperscript{th} edge and $i$\textsuperscript{th} node, respectively.
Plugging the above grid charges and currents to the discrete counterpart of continuity equation, one can prove the charge conservation \cite{moon2015exact}. 
Then, $\mathbf{j}^{n+\frac{1}{2}}$, which is a column vector including DoFs of grid currents, enters to the right hand side of (\ref{eq:dal_prime}) multiplied by $-\Delta t$ for field updates.
}
\clearpage
\section*{References}
\bibliography{mybibfile}

\end{document}